\documentclass[aps,prb,10pt,twocolumn,superscriptaddress,amsmath,amssymb,longbibliography,floatfix]{revtex4-2}
\usepackage[english]{babel}
\usepackage{graphicx,bbm}
\usepackage[utf8]{inputenc}
\usepackage[colorlinks,linkcolor=blue,citecolor=blue,urlcolor=blue]{hyperref}
\usepackage{xfrac}
\usepackage{soul}

\begin{document} 
\title{Magnon damping and mode softening in quantum double-exchange ferromagnets}
\author{A. Moreo}
\author{E. Dagotto}
\affiliation{Department of Physics and Astronomy, University of Tennessee, Knoxville, Tennessee 37996, USA}
\affiliation{Materials Science and Technology Division, Oak Ridge National Laboratory, Oak Ridge, Tennessee 37831, USA}
\author{G. Alvarez}
\affiliation{Computational Sciences and Engineering Division, Oak Ridge National Laboratory, Oak Ridge, Tennessee 37831, USA}
\author{T. Tohyama}
\affiliation{Department of Applied Physics, Tokyo University of Science, Tokyo 125-8585, Japan}
\author{M. Mierzejewski}
\author{J. Herbrych}
\affiliation{Institute of Theoretical Physics, Faculty of Fundamental Problems of Technology, Wroc{\l}aw University of Science and Technology, 50-370 Wroc{\l}aw, Poland}
\date{\today}
\begin{abstract}
We present a comprehensive analysis of the magnetic excitations and electronic properties of {\it fully quantum} double-exchange ferromagnets, i.e., systems where ferromagnetic ordering emerges from the competition between spin, charge, and orbital degrees of freedom, but without the canonical approximation of using classical localized spins. Specifically, we investigate spin excitations within the Kondo lattice-like model, as well as a two-orbital Hubbard Hamiltonian in proximity to the orbital-selective Mott phase. Computational analysis of the magnon dispersion, damping, and spectral weight within these models reveals unexpected phenomena, such as magnon mode softening and the anomalous decoherence of magnetic excitations as observed in earlier experimental efforts, but explained here without the use of the phononic degrees of freedom. We show that these effects are intrinsically linked to incoherent spectral features near the Fermi level, which arise due to the quantum nature of the local (on-site) triplets. This incoherent spectrum leads to a Stoner-like continuum on which spin excitations scatter, governing magnon lifetime and strongly influencing the dynamical spin structure factor. Our study explores the transition from coherent to incoherent magnon spectra by varying the electron density. Furthermore, we demonstrate that the magnitude of the localized spin mitigates decoherence by suppressing the incoherent spectral contributions near the Fermi level. We also discuss the effective $J_1$-$J_2$ spin Hamiltonian, which can accurately describe the large doping region characterized by the magnon-mode softening. Finally, we show that this behavior is also present in multiorbital models with partially filled orbitals, namely, in systems without localized spin moments, provided that the model is in a strong coupling regime. Our results potentially have far-reaching implications for understanding ferromagnetic ordering in various multi-band systems. These findings establish a previously unknown direct connection between the electronic correlations of those materials and spin excitations.
\end{abstract} 

\maketitle

\section{Introduction}
The ferromagnetism of transition metal materials remains a challenge despite nearly a century of investigation. Typically, one of three approaches is used to describe the magnetic properties of a given compound. (i) In systems with localized charge carriers, the localized magnetic moments form a lattice (i.e., Heisenberg spin model) and interact via an exchange mechanism \cite{Heisenberg1928}.  (ii) In contrast, delocalized Bloch plane waves mediate exchange interactions between spins in itinerant electron systems \cite{Moriya1979}. In such a situation, the scattering between electrons and holes gives rise to the so-called Stoner continuum~\cite{Stoner1947}, which influences the stability of the magnetic excitations. (iii) In the third scenario, both localized spins and itinerant electrons coexist. Depending on the hybridization between the latter and the strength of the interaction, the behavior of such systems is encapsulated in the Hubbard-Kanamori model \cite{Hubbard1963,Kanamori1963}, the periodic Anderson model \cite{Anderson1961}, or (derived in the limit of strong interactions \cite{Sinjukow2002}) in the Kondo-lattice model.

The third scenario is usually the right approach in strongly correlated systems with more than one valence band contributing to the Fermi level (i.e., in multiorbital systems). The textbook \cite{Kaplan1999,Dagotto2003,Furukawa2004} example of this phenomenon is present in transition metals with partially filled $d$ electron orbitals, with perovskite manganese oxides (manganites; R$_{1-x}$A$_x$MnO$_3$ where R=La,Ho,Nd,Pr and A=Sr,Ca,Pb) as a prime example. Here, three of four $3d$ electrons of Mn$^{3+}$ ions occupy $t_{2g}$ orbitals, and the remaining itinerant electron occupies one of the $e_g$ orbitals \cite{Dagotto2001}. The former are localized and form (due to the Hund rules) an effective magnetic moment of $S=3/2$ - often approximated by the semiclassical $S\to\infty$ limit because of its large value. Since manganites show a huge decrease in resistance by applying a magnetic field (the so-called colossal magnetoresistance) \cite{Fontcuberta1996,Dagotto2001}, considerable effort was devoted to describing the nontrivial physics of interplay between mobile electrons and localized spins. Consequently, the Kondo-lattice was extensively investigated in the past \cite{Yunoki2000,Hotta2000b,Hotta2001,Dagotto2001,Hotta2003,Dagotto2008,Luo2010}.

The magnetic properties of manganites and related compounds are strongly influenced by the double-exchange mechanism \cite{Zener1951,Anderson1955,deGennes1960}. In the generic scenario, two ions with different oxidation (e.g, Mn$^{3+}$ and Mn$^{4+}$ bridged by O$^{2-}$) can easily exchange the $e_g$ electron if its spin projection is ferromagnetically (FM) aligned with the remaining $t_{2g}$ ones. This constraint is a consequence of the ferromagnetic Hund exchange $J_\mathrm{H}$ present in the system. Due to this mechanism, double-exchange ferromagnetic ordering naturally occurs in multiorbital systems at electronic densities away from half-filling ($n\ne1$). Consider itinerant electrons interacting with localized magnetic moments. The interorbital Hund exchange $J_\mathrm{H}$ favors parallel alignment of their spins, forming maximized local spins. Let us discuss here a localized $S=1/2$ spin and only one itinerant orbital for simplicity. At half-filling $n=1$, states with maximum local spin (forming triplets in this case) are favored over doubly occupied or empty sites. Such system orders antiferromagnetically (AFM) via a superexchange mechanism with coupling $\propto 1/J_\mathrm{H}$, similar to the mechanism known from the single-orbital Hubbard model at large Hubbard interaction $U$. However, unlike one-band models, ferromagnetic ordering is preferred in multiorbital systems with large interactions ($J_\mathrm{H}\gg1$) and $n\ne1$. The latter emerges from the kinetic energy of the electrons. To minimize the energy, electrons that hop between neighboring sites must have the same spin projection as the localized spins. Consequently, the Hund interaction strongly couples electronic transport with the system's magnetism. This work will consider two models where this scenario occurs: the two-orbital Hubbard model and the generalized Kondo model (both described in the next section).

Although the ferromagnetic order is considered "trivial", with, e.g., simple Holstein–Primakoff magnons of \mbox{$[1-\cos(q)]$} dispersion, the excitations above double-exchange ferromagnets proved challenging. In the series of inelastic neutron scattering (INS) experiments on manganites \cite{Fernandez1998,Hwang1998,Vasiliu1998,Dai2000,Chatterji2002,Endoh2005,Ye2006,Zhang2007} unusual features, characteristic across the R$_{1-x}$A$_x$MnO$_3$ family, were found. For all densities $n$ the expected long-wavelength $q\to0$ quadratic Goldstone mode, $\omega(q)\propto q^2$, was observed. However, for shorter wavelengths, a sharp dispersion gave way to strong magnon decoherence and strongly $n$-dependent magnon mode softening. In the naive classical spin-wave consideration \cite{Ye2006,Zhang2007}, the latter required nearest-neighbor and not well-justified fourth-neighbor coupling (with vanishing second- and third-nearest-neighbor coupling) to reproduce the experimental dispersion. Various origins and approaches have been proposed to explain this behavior. Although the first-principle calculations can provide at least a qualitative description of "simple" metallic and half-metallic ferromagnets \cite{Gyorffy1985,Halilov1997,Irkhin1994,Sabiryanov1999,Coey2003,Katsnelson2008} it was suggested \cite{Oles1984,Filippetti2001,Medvedeva2001,Medvedeva2002} that Coulombic interaction might be important and effects beyond {\it ab initio} methods have to be considered. Other approaches included, e.g., phase separation and the presence of magnetic polarons \cite{Hotta2000,Dagotto2001,Koller2003,Neuber2006}, breakdown of the canonical double-exchange limit \cite{Solovyev1999}, non-Stoner continuum \cite{Kaplan2001}, spin-wave and $1/S$-expansion \cite{Wang1998,Vogt2001,Shannon2002,Schwabe2009,Frakulla2024}, and strong spin–lattice/orbital coupling \cite{Furukawa1999,Hotta1999,Hotta2000,Khaliullin2000,Hotta2001,Woods2001,Krivenko2004,Endoh2005}. Importantly, the latter is consistent with the experimental finding \cite{Fernandez1998,Dai2000,Ye2006,Ye2007} of the importance of Jahn–Teller phonons. However, as shown below, in the full quantum model (i.e., with $S=1/2$ localized moments), the unusual features of excitations above double-exchange ferromagnets are reproduced with the "simple" Kondo-like model without Jahn–Teller distortion.

In the context of multiorbital systems, ferromagnetic ordering also appears in other materials. Recently, much interest has been dedicated to the orbital-selective Mott phase in iron pnictides or ruthenates in which the ferromagnetically ordered phase can appear \cite{Koster2012,Georges2013,Fernandes2016,Mancini2017}, and to the coexistence of ferromagnetism and superconductivity in heavy-fermion materials \cite{Dikin2011,Bao2022,Wu2024}. An example of the latter is the exciting discovery \cite{Ran2019} of spin-triplet superconductivity in the U(Te,Ge)$_{2}$ family of materials. Finally, it is worth noting that non-equilibrium setups (i.e., pump-probe spectroscopy) can also induce a state with coexistent ferromagnetic and superconducting order in multiorbital Hubbard-Kanamori models \cite{Atsushi2018,Sujay2024}. In this context, we will show that even in systems without localized spin moments, the overall behavior is akin to the one described in the Kondo-lattice considerations.

To end this section let us list the main achievements of our work:
\begin{itemize}
\item[(i)] We present the magnon dispersion relation for a fully quantum double-exchange ferromagnet.
\item[(ii)] We demonstrate that the characteristic features of spin excitations identified experimentally in manganites — specifically, magnon mode softening and anomalous magnon damping — are present in the full many-body calculations of the Kondo-lattice model without phononic degrees of freedom.
\item[(iii)] We show that the single-particle spectral function of double-exchange ferromagnets includes incoherent spectral weight near the Fermi level. Such excitations control the magnon damping via scattering with a Stoner-like continuum. Since the latter emerges from the local multiplets present in the fully quantum system, they are not possible when only classical localized spins are considered and are discussed here.
\item[(iv)] We investigate how the magnon dispersion depends on the localized spin $S$ length, rendering our results relevant to many itinerant ferromagnets.
\item[(v)] We show that in the large doping region, $n\gtrsim1.6$, the spin excitations can be effectively described by the $J_1$-$J_2$ spin model. Our results confirm the phenomenological (experimental) observation that magnon mode softening has to be described with the help of second-nearest-neighbor interaction along the primary lattice directions in the effective spin model consideration.
\item[(vi)] We show that all of the above phenomena are also present in the full two-orbital model, i.e., in the parameter region where localized spins are absent.
\end{itemize}

The paper is structured as follows. In Sec.~\ref{sec:mome}, we set the stage for the discussion about excitations above the double-exchange ferromagnet. First, we examine the two-orbital Hubbard Hamiltonian and briefly describe how the orbital-selective Mott phase emerges in this model. Next, we present the effective description, namely the generalized Kondo model, for which most quantities will be evaluated. We conclude this section by discussing the methods and parameters used in this work. Sec.~\ref{sec:charge} contains the analysis of the single-particle spectral function and the density of states. Finally, in Sec.~\ref{sec:spin}, we present the main results: the spin excitations analysis of the quantum double-exchange ferromagnet. This section also describes the Stoner-like continuum, which is necessary for understanding spin excitations, and also the effective long-range spin Hamiltonian. Sec.~\ref{sec:spintwoorb} is devoted to spin excitations in the two-orbital Hubbard model. The discussion and conclusions are given in Sec.~\ref{sec:concl}. In the Appendix, we present additional results for the charge structure factor (App.~\ref{app:charge}) and the dependence on the system parameters (App.~\ref{app:para} and App.~\ref{app:fit}).

\section{Models and methods}
\label{sec:mome}

\subsection*{Two-orbital Hubbard-Kanamori model \& orbital-selective Mott phase}
\label{sec:to}

To be practical, we will carry out our study in one dimension (1D) because the results are quasi-exact by using powerful computational techniques. However, we argue that our conclusions are generic and simple and could apply to higher dimensions as well. The two-orbital 1D Hubbard-Kanamori model (HK) is given by
\begin{eqnarray}
H_\mathrm{2O}&=& \sum_{\gamma\gamma^\prime\ell\sigma} t_{\gamma\gamma^\prime}
\left(c^{\dagger}_{\gamma\ell\sigma}c^{\phantom{\dagger}}_{\gamma^\prime\ell+1\sigma}+\mathrm{H.c.}\right)+\Delta_\mathrm{CF}\sum_\ell n_{1\ell}\nonumber\\
&+& U\sum_{\gamma\ell}n_{\gamma\ell\uparrow}n_{\gamma\ell\downarrow}
+U^\prime \sum_{\ell} n_{0\ell} n_{1\ell}\nonumber\\
&-& 2J_\mathrm{H} \sum_{\ell} \mathbf{S}_{0\ell} \cdot \mathbf{S}_{1\ell}
+J_\mathrm{H} \sum_{\ell} \left(P^{\dagger}_{0\ell}P^{\phantom{\dagger}}_{1\ell}+\mathrm{H.c.}\right)\,.
\label{eq:ham2o}
\end{eqnarray}
Here $c^\dagger_{\gamma\ell\sigma}$ ($c_{\gamma\ell\sigma}$) represent electron creation (annihilation) operator at orbital $\gamma=\{0,1\}$ and site $\ell=\{1,\dots,L\}$, $n_{\gamma\ell\sigma}$ represents density operator (with $n_{\gamma\ell}=n_{\gamma\ell\uparrow}+n_{\gamma\ell\downarrow}$), $S_{\gamma\ell}$ is the local spin, and $P^\dagger_{\gamma\ell}=c^\dagger_{\gamma\uparrow\ell}c^\dagger_{\gamma\downarrow\ell}$. The first two terms in the above equation represent the system’s kinetic energy, with $t$ as hopping and $\Delta_\mathrm{CF}$ as the crystal field, respectively. The rest of the Hamiltonian accounts for potential energy: the third and fourth terms describe the on-site electron repulsion, with intra-orbital $U$ and inter-orbital $U^\prime$ interactions. The last two terms originate from the multiorbital physics: $J_\mathrm{H}$ accounts for the ferromagnetic Hund coupling between spins $\mathbf{S}_{\gamma \ell}$ in different orbitals, maximizing the total on-site spin. The model preserves SU(2) symmetry when $U^\prime =U-5/2J_\mathrm{H}$ \cite{Georges2013}. 

The phase diagram of~\eqref{eq:ham2o}, i.e., as a function of the interaction $U$  strongly depends on the value of the Hund exchange $J_\mathrm{H}$. At $J_\mathrm{H}\to0$, one can observe the ``standard" metal to Mott insulator transition, familiar from the single-orbital Hubbard model. For $J_\mathrm{H}\ne0$, especially when $J_\mathrm{H}\sim U$, a new phase can emerge \cite{Vojta2010,Georges2013}. In the case of orbital differentiation (e.g., $t_{00}>t_{11}$, or/and due to the presence of a crystal field $\Delta_\mathrm{CF}\ne0$), one of the orbitals can undergo a Mott phase transition, while the other remains itinerant (metallic). This phenomenon is termed the orbital-selective Mott phase (OSMP). Consequently, within the OSMP, itinerant electrons interact with localized ones. Although the exact conditions under which the OSMP is stabilized are still under study \cite{Kugler2022,Stepanov2022,Hu2022}, this phase appears to be relevant for various iron pnictides and/or ruthenates \cite{Georges2013,Fernandes2016,Mancini2017}, including systems with reduced dimensionality \cite{Rincon2014,Herbrych2018,Sroda2021,Lin2022}.

The OSMP, i.e., the unique "mixture" of metallic and insulating bands, leads to a rich magnetic phase diagram: (i) At small interaction values, the paramagnetic state dominates. (ii) At $U\sim J_\mathrm{H}\sim W$, where $W$ is the kinetic energy bandwidth, the system is in a block-magnetic state \cite{Mourigal2015,Rincon2014,Herbrych2018,Herbrych2019,Herbrych2020,Sroda2021}, i.e., AFM coupled FM islands (blocks) of various $n$-dependent sizes. (iii) For $J_\mathrm{H}\gg t$ and $n\ne 1$ eventually the expected fully ferromagnetic ordering is present. This is exactly the region of interest of this work. (iv) Finally, between FM and block phases, an incommensurate block-spiral ordering can be stabilized \cite{Herbrych2020-2} due to the competition between FM (due to double exchange) and AFM (due to superexchange) tendencies.

\subsection*{Generalized Kondo model}

It has been shown \cite{Herbrych2019,Herbrych2020,Herbrych2020-2} that the static and dynamic quantities within the OMSP can be very accurately described by the generalized Kondo (gK) model, given by:
\begin{eqnarray}
H_\mathrm{gK} &=& t\sum_{\ell\sigma}
\left(c^{\dagger}_{\ell\sigma}c^{\phantom{\dagger}}_{\ell+1\sigma}+\mathrm{H.c.}\right)
+U\sum_{\ell}n_{\ell\uparrow}n_{\ell\downarrow}\nonumber\\
&-&2J_\mathrm{H}\sum_{\ell}\mathbf{s}_{\ell} \cdot \mathbf{S}_{\ell}\,,
\label{eq:hamkh}
\end{eqnarray}
with $t=t_{00}$, provided that $t_{00}$ represent the hopping of wide band ($t_{00}\gg t_{11}$). Similarly, $\mathbf{s}_{\ell}=\mathbf{S}_{0\ell}$ and $\mathbf{S}_{\ell}=\mathbf{S}_{1\ell}$ represent the spin of an electron at the wide and narrow bands, respectively. Note that for half-filling $n=1$ and in the $U\,,J_\mathrm{H}\gg t$ limit, the total magnetic moment squared 
\begin{equation}
T^2=T(T+1)\,,\quad T=s+S=1/2+S\,,
\end{equation}
is maximized, e.g., $T=1$ for $S=1/2$ localized spins. The total magnetization \mbox{$T^z_\mathrm{tot}=s^z_\mathrm{tot}+S^z_\mathrm{tot}$} is conserved, $[H,T^z_\mathrm{tot}]=0$ while the band magnetizations are not, $[H,s^z_\mathrm{tot}]\ne0\ne[H,S^z_\mathrm{tot}]$. 

As described in the introduction, semiclassical versions of Kondo lattice models (with $U=0$ and $S\to\infty$) have been extensively studied in the past. This approach is justified for manganese oxides, where the $J_\mathrm{H}|\mathbf{S}|\gg1$ limit of~\eqref{eq:hamkh} is believed to provide at least qualitatively the correct result. On the other hand, for compounds where the OSMP is relevant, only one orbital is Mott localized with $S=1/2$. In this case, one must consider the full quantum version of gK since $J_\mathrm{H}|\mathbf{S}|\sim U\sim W$. In this work, we will predominantly focus on the $S=1/2$ case; however, we will also comment on the dependence of our findings on $S$, including discussions for the cases of $S=1$ and $S=3/2$ localized spins coupled to $s=1/2$ itinerant fermions. The latter can be viewed as a particular case of the three- and four-orbital Hubbard model~\eqref{eq:ham2o}, respectively, with only one itinerant band.

\subsection*{Methods}

We will primarily focus on two experimentally relevant quantities: the single-particle spectral function $A(q,\omega)$ (discussed in Sec.~\ref{sec:charge}) and the dynamical spin structure factor $S(q,\omega)$ (discussed in Sec.~\ref{sec:spin}). The former is directly relevant for angle-resolved photoemission spectroscopy (ARPES) \cite{Damascelli2004,Lu2012,Sobota2021,Zhang2022,Boschini2024}, while the latter pertains to INS \cite{Lovesey1984,Squires2012,Fujita2012,Dai2015}. Our results are also relevant for resonant inelastic X-ray scattering (RIXS) \cite{Kotani2001,Ament2011,Mitrano2024}. Below, we discuss the excitations above the ground state, i.e., we will consider the zero temperature limit. Both of these dynamical (frequency-dependent) quantities can be expressed as the imaginary part of Green's functions of the form
\begin{equation}
\langle\langle A_{\ell}\,B_{m} \rangle\rangle_{\omega}^{\mp}=-\frac{1}{\pi}\mathrm{Im}\langle \mathrm{gs}|A_{\ell}\frac{1}{\omega^{+}\mp(H-\epsilon_\mathrm{GS})}B_{m}|\mathrm{gs}\rangle\,,
\label{eq:green}
\end{equation}
where $\omega^{+}=\omega+i\eta$ with $\eta$ as an internal broadening (caused by secondary effects such as disorder or resolution), $|\mathrm{gs}\rangle$ represents the ground state vector with energy $\epsilon_\mathrm{GS}$, and $A_l\,,B_m$ are appropriate operators. Finally, in the following we will consider Green's functions in the momentum space obtained from the Fourier transform (FT) of Eq.~\eqref{eq:green}, e.g., $f(q,\omega)\propto \sum_{\ell}\mathrm{e}^{i(\ell-L/2)q}\langle\langle A_{\ell}\,B_{m} \rangle\rangle_{\omega}^{\mp}$. Note that in order to decrees computational resources, we have changed the double to single sum in FT, i.e., we have fixed the second index to center of the chain $m=L/2$. This is a standard procedure \cite{Kuhner1999} for a open-boundary systems consider in this work. In the latter, a possible artifact of such FT is systematic "beating" with a small amplitude visible, e.g., in Fig.~\ref{fig:akw}. In our analysis we made sure that this is indeed just a spurious feature, e.g., see detail analysis presented in Fig.~\ref{fig:akw_polarized}.

The ground states of the Hamiltonians and the Green's functions analyzed in this work were obtained using the density matrix renormalization group (DMRG) method~\cite{White1992,Schollwock2005} within the single-site approach~\cite{White2005}. Dynamical correlation functions were computed via the dynamical-DMRG~\cite{Jeckelmann2002,Nocera2016}, using the correction-vector method and Krylov decomposition to calculate spectral functions in frequency space~\cite{Nocera2016} directly. During the DMRG simulations, up to $M=1536$ states were retained, enabling us to simulate systems as large as $L=200$ for the gK model~\eqref{eq:hamkh} and $L=60$ for the two-orbital Hubbard model~\eqref{eq:ham2o}, with truncation errors below $10^{-7}$.

If not otherwise stated, in the main part of our work (i.e., in Sec.~\ref{sec:charge} and Sec.~\ref{sec:spin}), we will use the gK model~\eqref{eq:hamkh} in the large Hund limit, i.e., with $t=0.5\,\mathrm{[eV]}$, $U/t=20$, $J_\mathrm{H}/t=20$, $L=200$ sites, and $S=1/2$ localized spins. Although $U=J_\mathrm{H}$ would be considered reasonable to describe manganites phenomenologically, in the last part of the publication, we relax these numbers towards more realistic values. In Sec.~\ref{sec:spinlen}, we will consider $S=1$ and $S=3/2$, while additional results for various system parameters ($U/t=0,10$ and $J_\mathrm{H}/t=10,40$) are given in App.~\ref{app:para}. Finally, the results for the full two-orbital Hubbard model~\eqref{eq:ham2o} will be presented in Sec.~\ref{sec:spintwoorb}.

\section{Electronic excitations}
\label{sec:charge}
Let us first focus on the single-particle spectral function, which is defined as
\begin{eqnarray}
A(q,\omega)=\frac{1}{L}\sum_{\ell} \mathrm{e}^{i (\ell-L/2) q}\,\left(
\langle\langle c^{\phantom{\dagger}}_{\ell} c^{\dagger}_{L/2} \rangle\rangle_{\omega}^{-}+
\langle\langle c^{\dagger}_{\ell} c^{\phantom{\dagger}}_{L/2} \rangle\rangle_{\omega}^{+}
\right)\,,
\label{eq:akw}
\end{eqnarray}
with $c_{\ell}=\sum_\sigma c_{\ell\sigma}$ and $\langle\langle c^{\dagger}_{\ell} c^{\phantom{\dagger}}_{L/2} \rangle\rangle_{\omega}^{+}$ ($\langle\langle c^{\phantom{\dagger}}_{\ell} c^{\dagger}_{L/2} \rangle\rangle_{\omega}^{-}$) representing the hole (electron) dynamics below (above) the Fermi level $\epsilon_\mathrm{F}$. We will also consider the spin-resolved spectral functions $A_\sigma(q,\omega)$, where $\sigma=\uparrow,\downarrow$ denotes the spin of the creation/annihilation operators [i.e., $c_{\ell\sigma}$ instead of $c_{\ell}$ in Eq.~\eqref{eq:akw}]. Note that $A(q,\omega)=A_\uparrow(q,\omega)+A_\downarrow(q,\omega)$ due to the spin conservation of $H_\mathrm{gK}$~\eqref{eq:hamkh}. In the remainder of this Section (if not otherwise stated), we will consider $H_\mathrm{gK}-\mu N$, with the chemical potential $\mu$ set to the Fermi level ($\mu=\epsilon_\mathrm{F}$). With this definition, the $\omega>0$ ($\omega<0$) results represent the electron (hole) part of the spectral function.

\begin{figure*}[!htb]
\includegraphics[width=1.0\textwidth]{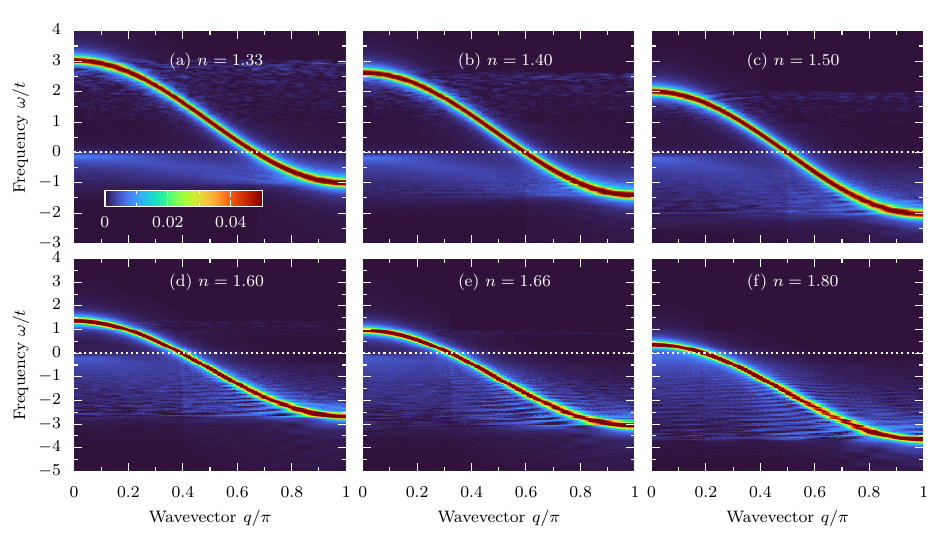}
\caption{Single-particle spectral function $A(q,\omega)$ of the generalized Kondo (gK) model with $S=1/2$ localized spins in the limit of $J_\mathrm{H}\gg t$ ($U/t=J_\mathrm{H}/t=20$ and $T^z_\mathrm{tot}=0$ magnetization sector) for various electron doping levels, $n=1.33\,,1.40\,,1.50\,,1.60\,,1.66\,,1.80$ [panels (a) to (f), respectively]. Note the visible spectral weight below the Fermi level $\omega=0$ (the latter is depicted as a white dashed line). In all panels: $L=200$ sites, frequency resolution $\delta\omega/t=4\cdot 10^{-2}$, and $\eta=2\delta\omega$.}
\label{fig:akw}
\end{figure*}

In Fig.~\ref{fig:akw}, we present the electron density $n>1$ dependence of the single-particle spectral function $A(q,\omega)$ in proximity to the Fermi level for the gK model~\eqref{eq:hamkh}. Due to the particle-hole symmetry of the latter, our results are equivalent to $2-n$ filling. The parameters used here are: $S=1/2$ localized spins, \mbox{$U/t=J_\mathrm{H}/t=20$}, and zero magnetization sector $T^z_\mathrm{tot}=0$. It is important to note that additional bands of excitations are present deep below the Fermi level. We will briefly comment on them later and refer the interested reader to Ref.~\cite{Sroda2023} for a detailed discussion.

The results presented in Fig.~\ref{fig:akw} show that the coherent spectrum in proximity to the Fermi level resembles that of noninteracting spinless fermions and can be modeled with
\begin{equation}
\omega_\mathrm{ff}(q)=2t\cos(q)-\mu_\mathrm{ff}\,,
\label{eq:ff}
\end{equation}
where $\mu_\mathrm{ff}=2t\cos\left(k^\mathrm{ff}_\mathrm{F}\right)$ and $k^\mathrm{ff}_\mathrm{F}=\pi (2-n)$, i.e., by the system of noninteracting spinless fermions with effective density $n_\mathrm{ff}=2-n$. Note, however, the different Fermi level dependence on the density $n$, given by $2k_\mathrm{F}=\pi n$, for the noninteracting spinfull electrons [i.e., for the \mbox{$U\to0\,,J_\mathrm{H}\to0$} limit of the gK model \eqref{eq:hamkh} considered here]. Such findings are consistent with the semiclassical result in the $S\to\infty$ limit~\cite{Yunoki1998}.

The above behavior can be easily understood in the ferromagnetically polarized system (i.e., for $n\ne1$ and $J_\mathrm{H}\gg t$). Consider the density $n=1.5$, whose spectral function resembles the half-filled free fermion case; see Fig.~\ref{fig:akw}(c). To minimize the kinetic energy, the ground state is built from an equal proportion of singlons/triplets and doublons (eigenstates of the atomic limit, $t\to0$),
\begin{equation*}
|\mathrm{gs}\rangle_\mathrm{atomic}=
\frac{1}{\sqrt{2}}
\left|
\begin{matrix}
\uparrow\\
2
\end{matrix}
\right\rangle
-
\frac{1}{\sqrt{2}}\left|
\begin{matrix}
\uparrow\\
\uparrow
\end{matrix}
\right\rangle
\,,
\end{equation*}
arranged in a staggered fashion to maximize the mobility of the electrons. We can pictorially represent such states as empty and occupied sites of a spinless fermions-like system, i.e.,
\begin{equation*}
\left|
\begin{matrix}
\uparrow & \uparrow & \uparrow & \uparrow \\
2        & \uparrow & 2        & \uparrow
\end{matrix}
\right\rangle
\simeq
|1010\rangle\,.
\end{equation*}
Here, the state on the left depicts a sketch of the ground state of a Kondo-like model with all localized spins polarized (up row) and $n=1.5$ electrons in the itinerant band (bottom row). Note that the true many-body ground state is a quantum liquid built predominantly from the configurations of the above type. Even in the case of the polarized state, \mbox{$T^z_\mathrm{tot}=SL+sL(2-n)$}, the electrons in the itinerant orbital don't form any apparent CDW (provided that the system is not in the phase-separated state, which don't discuss in this work).

\begin{figure}[!htb]
\includegraphics[width=1.0\columnwidth]{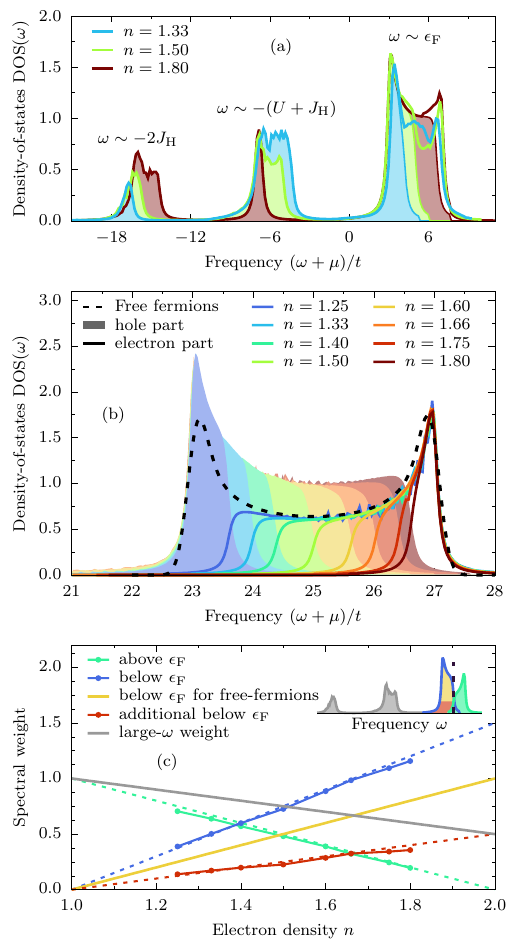}
\caption{(a) Density-of-states $\mathrm{DOS}(\omega)$ of the gK model with $S=1/2$ localized spins in the $J_\mathrm{H}\gg t$ limit. Three separate bands are clearly visible (at $\omega\sim \epsilon_\mathrm{F}$, $\omega\sim -2J_\mathrm{H}$, and $\omega\sim -U-J_\mathrm{H}$). Evaluated for $L=200$, $U=0$, $J_\mathrm{H}/t=20$, and $T^z_\mathrm{tot}=0$ magnetization sector. (b) $\mathrm{DOS}(\omega)$ close to the Fermi level (obtained from the $q$-integrating data presented in Fig.~\ref{fig:akw}, i.e., evaluated for $L=200$, $U/t=J_\mathrm{H}/t=20$, and $T^z_\mathrm{tot}=0$). Solid colors depict spectral weight below the Fermi level (hole part $\propto \langle\langle c^{\dagger}_{\ell} c^{\phantom{\dagger}}_{L/2} \rangle\rangle_{\omega}^{+}$). Lines depict results above the Fermi level (electron part $\propto \langle\langle c^{\phantom{\dagger}}_{\ell} c^{\dagger}_{L/2} \rangle\rangle_{\omega}^{-})$. The free-fermion solution is shown as a dashed line. As evident from the presented results, below the Fermi level $\mu=\epsilon_\mathrm{F}$, one can observe additional spectral weight (w.r.t. the free fermions solution). (c) Integrated spectral weight $N_\mathrm{F}$ of the data in panel (b). The colored dashed lines represent guides to the eye. The inset depicts the frequency range of a given spectral weight. See the text for details.}
\label{fig:dos}
\end{figure}

With this naive mapping to spinless fermions, the $2t\cos(q)$ dispersion for $\omega>0$ indicates that the noninteracting considerations could describe the single-particle excitations near the Fermi level. However, this simple picture fails to capture important (from the perspective of magnetic excitations) details of the $\omega<0$ spectrum. This becomes evident from the analysis of the density-of-states (DOS) given by $\mathrm{DOS}(\omega)=\sum_q\,A(q,\omega)$. In Fig.~\ref{fig:dos}(a), we present the complete DOS over the wide range of frequencies, while in Fig.~\ref{fig:dos}(b), we provide detailed results in the proximity of the Fermi level. In order to visualize non-overlapping bands in the spectrum, in Fig.~\ref{fig:dos}(a), we use $U=0$ and $J_\mathrm{H}/t=20$. In addition to the spectral weight close to the Fermi level, two additional bands can be found: (i) at $\omega\sim -2J_\mathrm{H}$ the local (on-site) triplet to local singlet excitations, known as Hund band excitations, which breaks the Hund's rules \cite{Bauernfeind2017,Stadler2019,Sroda2023}; and (ii) at $\omega\sim-(U+J_\mathrm{H})$, the "standard" singlon-doublon excitation known from the single-orbital Hubbard model (shifted additionally by $J_\mathrm{H}$). 

Let us focus on the spectral weight near the Fermi energy. In Fig.~\ref{fig:dos}(b), we present the DOS corresponding to the data in Fig.~\ref{fig:akw} (i.e., $U/t=J_\mathrm{H}/t=20$). For all considered electron densities $n$, the $\mathrm{DOS}(\omega>0)$ is perfectly reproduced by the noninteracting solution \mbox{$\mathrm{DOS}_\mathrm{ff}(\omega)=1/2\pi\sin[\arccos(\omega/2)]$} [indicated as the black dashed line in Fig.~\ref{fig:dos}(b)]. On the other hand, for $\omega<0$, we find additional spectral weight. This can be quantified by integrating the spectral weight of the band in the neighborhood to the Fermi level: $N_{F}=\int\mathrm{d}\omega\,\mathrm{DOS}(\omega)$ [see Fig.~\ref{fig:dos}(c)]. Consider again the case of $n=1.5$. Taking the $\omega>0$ weight [green line in Fig.~\ref{fig:dos}(c)] as a reference point — i.e., the part of the spectrum perfectly described by the free fermion solution at half-filling [see Fig.~\ref{fig:akw}(c)] — we would expect the same weight for $\omega<0$ (yellow line). Surprisingly, our results indicate a $1.5$ times larger contribution for all considered $n$ (blue and red lines).

\begin{figure*}[!htb]
\includegraphics[width=1.0\textwidth]{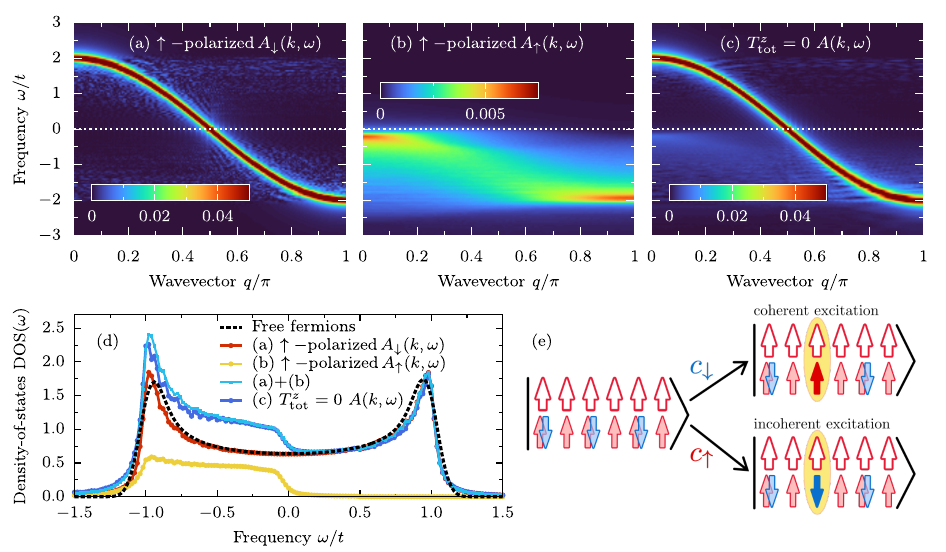}
\caption{Single-particle spectral function $A(q,\omega)$ of the $\uparrow$-polarized gK model with $S=1/2$ localized spins, calculated for $n=1.5$, $U/t=J_\mathrm{H}/t=20$, and the $T^z_\mathrm{tot}=SL+sL(2-n)$ magnetization sector. (a,b) Spin-resolved $A_\sigma(q,\omega)$ with (a) a spin projection antiparallel $\sigma=\downarrow$ (coherent band) and (b) parallel $\sigma=\uparrow$ (incoherent band) to the polarization of the system. Note the change in color scale. Panel (c) depicts the spectral function for the $T^z_\mathrm{tot}=0$ case. (d) The density-of-states $\mathrm{DOS}(\omega)$ of the data presented in panels (a-c). It is evident that the DOS of the antiparallel case [panel (a)] is perfectly described by the free fermion solution (indicated by the black dashed line). (e) Sketch of the coherent and incoherent excitations. In all panels: $L=200$, $\delta\omega/t=4\cdot 10^{-2}$ and $\eta=2\delta\omega$.}
\label{fig:akw_polarized}
\end{figure*}

To gain further insight into the spectrum below the Fermi level, let us focus on the fully up-polarized system, i.e., \mbox{$T^z_\mathrm{tot}=SL+sL(2-n)$} for $n>1$. Note that in the SU(2) symmetric system, the ferromagnetic ground state is not unique; all magnetization sectors are degenerate, and we expect the same behavior in all of them. Let's examine how the creation and annihilation operators act on the ground state. Consider operators with spin antiparallel to the polarization ($\sigma=\downarrow$). The creation operator $c^\dagger_{\ell\downarrow}$ can act on any of the local triplets, promoting it to double-occupied sites, e.g.,
\begin{equation*}
c^\dagger_{\ell=2\downarrow}\left|
\begin{matrix}
\uparrow & \uparrow & \uparrow & \uparrow & \uparrow & \uparrow\\
2        & \uparrow & 2        & \uparrow & 2        & \uparrow
\end{matrix}
\right\rangle
=
\left|
\begin{matrix}
\uparrow & \uparrow & \uparrow & \uparrow & \uparrow & \uparrow\\
2        & 2        & 2        & \uparrow & 2        & \uparrow
\end{matrix}
\right\rangle
\simeq
|111010\rangle\,,
\end{equation*}
here using $\ell=2$ as an example (we sketch only one of the possible spin-projections of the many-body state). Conversely, a down annihilation operator can promote any of the doublons to local triplets
\begin{equation*}
c_{\ell=3\downarrow}\left|
\begin{matrix}
\uparrow & \uparrow & \uparrow & \uparrow & \uparrow & \uparrow\\
2        & \uparrow & 2        & \uparrow & 2        & \uparrow
\end{matrix}
\right\rangle
=
\left|
\begin{matrix}
\uparrow & \uparrow & \uparrow & \uparrow & \uparrow & \uparrow\\
2        & \uparrow & \uparrow & \uparrow & 2        & \uparrow
\end{matrix}
\right\rangle
\simeq
|100010\rangle\,.
\end{equation*}
The free fermion-like considerations perfectly capture such processes. This is also reflected in the spin-resolved spectral function in the direction opposite to the magnetization, as shown in Fig.~\ref{fig:akw_polarized}(a), and the spin-resolved DOS in Fig.~\ref{fig:akw_polarized}(d). The $A_\downarrow(q,\omega)$ along with its $\mathrm{DOS}(\omega)$ are accurately represented by the noninteracting solution across the entire range of frequencies (near $\epsilon_\mathrm{F}$).

The action of the $c_{\ell\sigma}$/$c^\dagger_{\ell\sigma}$ operators with spin parallel to the polarization ($\sigma=\uparrow$) is different. First, no additional $\sigma=\uparrow$ electron can be created since the system is fully polarized. On the other hand, there are two possibilities for annihilating such electrons. Firstly, one can remove one electron from singly occupied sites, e.g.,
\begin{equation*}
c_{\ell=2\uparrow}\left|
\begin{matrix}
\uparrow & \uparrow & \uparrow & \uparrow & \uparrow & \uparrow\\
2        & \uparrow & 2        & \uparrow & 2        & \uparrow
\end{matrix}
\right\rangle
=
\left|
\begin{matrix}
\uparrow & \uparrow & \uparrow & \uparrow & \uparrow & \uparrow\\
2        & 0        & 2        & \uparrow & 2        & \uparrow
\end{matrix}
\right\rangle\,.
\end{equation*}
Such a state contributes to high energy excitations with $\omega\sim-(U+J_\mathrm{H})$. Secondly, the annihilation of an up electron from one of the doublons leads to the creation of a local antiparallel spin configuration
\begin{equation*}
c_{\ell=3\uparrow}\left|
\begin{matrix}
\uparrow & \uparrow & \uparrow & \uparrow & \uparrow & \uparrow\\
2        & \uparrow & 2        & \uparrow & 2        & \uparrow
\end{matrix}
\right\rangle
=
\left|
\begin{matrix}
\uparrow & \uparrow & \uparrow & \uparrow & \uparrow & \uparrow\\
2        & \uparrow & \downarrow & \uparrow & 2      & \uparrow
\end{matrix}
\right\rangle\,.
\end{equation*}
Both configurations obtained from the $c_{\ell\uparrow}$ action go beyond the simple free-fermion-like considerations, as such states cannot be mapped using the $|0\rangle$ and $|1\rangle$ states alone (representing double-occupied sites and local triplets, respectively, in our convention). Furthermore, the local antiparallel spin configuration is not an eigenstate of the model in the atomic limit. However, it has a finite projection onto eigenstates with local $T^z_\ell=0$, i.e., onto the singlet and one of the triplets. The former contributes to the high-frequency states, with \mbox{$\omega\propto -2J_\mathrm{H}$} because they violate Hund's rules. In contrast, the local triplet states form the ground state in the atomic limit ($t\to0$). In the many-body system ($t\ne0$), the action of the up annihilation operator on the up-polarized state creates an incoherent band of excitations below the Fermi level. This behavior is illustrated in Fig.~\ref{fig:akw_polarized}(b).

The remarkable picture emerging from our investigation indicates that a free-fermion-like solution only qualitatively describes the single-particle spectral function $A(q,\omega)$ of the Kondo-like model with $S=1/2$ localized spins. In the polarized system [$T^z_\mathrm{tot}=SL+sL(2-n)$], the dispersion of the removed electron (the hole part of the spectrum) depends on its spin. The electrons created or annihilated with opposite spin to the polarization form a coherent band perfectly described by a noninteracting solution. On the other hand, electrons annihilated with the same spin as the polarization form an incoherent band of excitations due to the projection to local triplet states that can build the ground state in the atomic limit (where the dimensionality does not play any role). As a consequence, we speculate that a similar incoherent band in $A(q,\omega)$ should be found in any lattice dimension since the above is a consequence of the quantum nature of the triplet (absent in the $S\to\infty$ limit due to lack of $T^z_\ell=0$ triplet projection).

\begin{figure*}[!htb]
\includegraphics[width=1.0\textwidth]{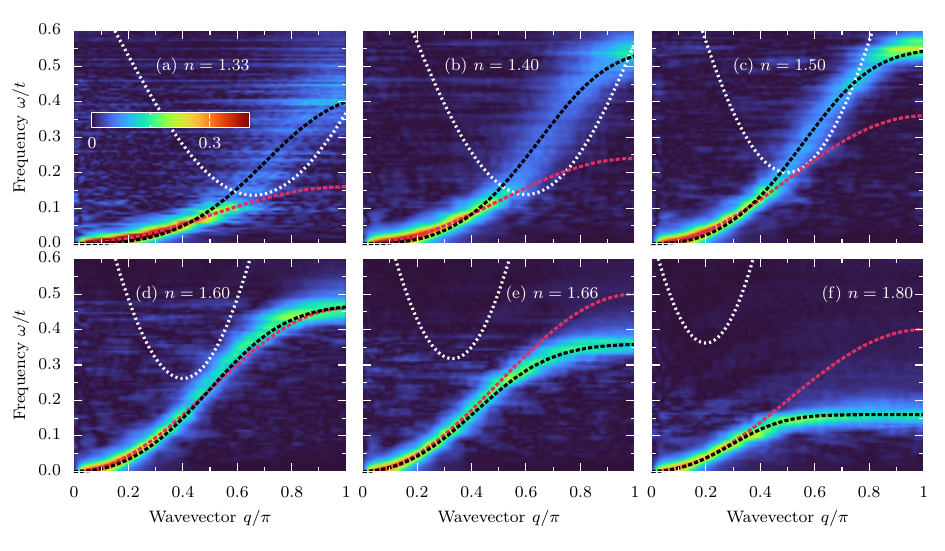}
\caption{Dynamical spin structure factor $S(q,\omega)$ of the gK model with $S=1/2$ localized spins in the $J_\mathrm{H}\gg t$ limit ($U/t=J_\mathrm{H}/t=20$, $T^z_\mathrm{tot}=0$ magnetization sector) for various electron doping levels, $n=1.33\,,1.40\,,1.50\,,1.60\,,1.66\,,1.80$ [panels (a) to (f), respectively]. Note the magnon damping (incoherent spectrum) for $n\lesssim 1.5$ and the magnon mode softening for $n\gtrsim 1.5$. The red dashed line represents the $\omega_\mathrm{m}(q)=J_{q\to0} [1-\cos(q)]$ coherent magnon dispersion with $J_{q\to0}$ obtained from the $q<\pi/4$ fit. The black dashed line represent fit to the $\omega_\mathrm{fit}(q)$ (see Sec.~\ref{sec:magsoft}). The white dashed line in all panels represents the bottom of the Stoner continuum obtained from the incoherent band (see text for details). In all panels: $L=200$, $\delta\omega/t=6\cdot 10^{-3}$ and $\eta=2\delta\omega$.}
\label{fig:sqw}
\end{figure*}

Note also that the total spectral function $A(q,\omega)$ is identical to the case of zero magnetization ($T^z_\mathrm{tot}=0$) [see Fig.~\ref{fig:akw_polarized}(c)]. However, in this case, both spin projections contribute to both bands due to the complicated nature of the many-body state at $T^z_\mathrm{tot}=0$. Nevertheless, an inspection of the single-particle spectral function $A(q,\omega)$ for zero magnetization, presented in Fig.~\ref{fig:akw}, indicates that an incoherent band of excitations is indeed present for all considered electron densities $n$ and spans from $\omega\simeq 0$ to $\omega\simeq\omega_\mathrm{ff}(\pi)$, i.e., from the Fermi level to the bottom of the noninteracting band. The analysis of the density-of-states [see Fig.~\ref{fig:dos} and Fig.~\ref{fig:akw_polarized}(d)] indicates that $1/3$ of spectral weight close to the Fermi level is in the incoherent band, with the remaining $2/3$ in the free-fermion-like dispersion $\sim\cos(q)$, irrespective of the magnetization of the system. Consequently, the additional weight in the $A(q,\omega)$ [or $\mathrm{DOS}(\omega)$] should be visible in photoemission ARPES experiments,  a novel prediction of our effort. Importantly, as shown in the next section, the incoherent band is necessary to understand the behavior of spin excitations.

\section{Spin excitations}
\label{sec:spin}

\begin{figure}[!htb]
\includegraphics[width=1.0\columnwidth]{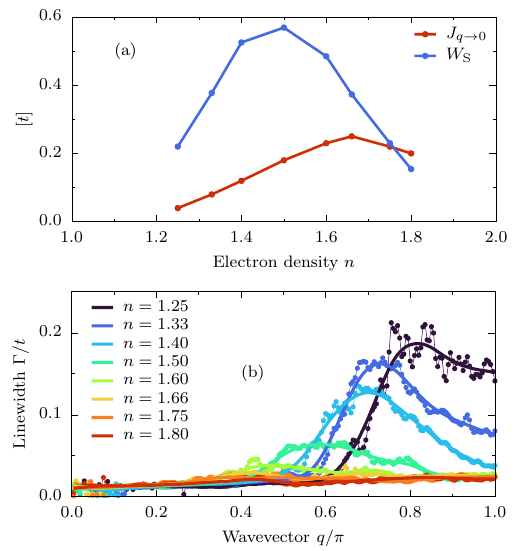}
\caption{(a) Dependence of the long-wavelength effective spin exchange $J_{q\to0}$ and the spin excitation bandwidth $W_\mathrm{S}$ on the electron density $n$. (b) Wavevector $q$ dependence of the magnon linewidth $\Gamma$, obtained from Lorentzian fits of the spin excitation spectrum shown in Fig.~\ref{fig:sqw} (see text for details). Points represent the result of the fit, Eq.~\eqref{eq:fit}, while solid lines represent guides to the eye.}
\label{fig:width}
\end{figure}

In this section, we will discuss the dispersion of spin excitations as measured by the dynamical spin structure factor
\begin{equation}
S(q,\omega)=\frac{1}{L}\sum_{\ell} \mathrm{e}^{i (\ell-L/2) q}\,
\langle\langle \mathbf{T}_{\ell}\mathbf{T}_{L/2} \rangle\rangle_{\omega}^{-}\,.
\label{eq:sqw}
\end{equation}
Here $\mathbf{T}_{\ell}=\mathbf{s}_{\ell}+\mathbf{S}_{\ell}$ is the total spin at site $\ell$. One of the main results of this work is presented in Fig.~\ref{fig:sqw}, which shows the electron density $n$ dependence of $S(q,\omega)$. Several conclusions can be drawn directly from the presented results. The quadratic behavior of long wavelength magnons can be clearly recognized for all considered cases, i.e., $\omega(q\to0)\propto q^2$. Fig.~\ref{fig:width}(a) depicts fits of the dispersion to $\omega_\mathrm{m}(q)=J_{q\to 0}q^2$ in the $0<q<\pi/4$ region. Our results indicate that the long-wavelength effective spin exchange $J_{q\to 0}$ increases till $n\simeq1.6$ and decreases afterward.

However, none of the considered cases can be fully described by the coherent magnon dispersion \mbox{$\omega_\mathrm{m}(q)=J_{q\to0} [1-\cos(q)]$}. The behavior at shorter wavelengths ($q/\pi\gtrsim0.4$) strongly depends on the doping. For $n>1.5$, we observe a gradual softening of the magnetic excitations with increasing electron density, along with a momentum-independent mode across a wide range of wavelengths $\pi/2<q<\pi$ for the largest considered density $n=1.8$. The behavior for $n\lesssim 1.5$ is strikingly different. Here, we note a highly incoherent dispersion for short wavelengths; namely, the magnons significantly reduce their lifetime $\Gamma$. In Fig.~\ref{fig:width}(b), we present the wavevector dependence of the magnon linewidths $\Gamma(q)$ obtained from the Lorentzian-like fits of $S(q,\omega)$ for a given $q$, i.e.,
\begin{equation}
f(\omega)=\frac{\Lambda}{(\omega-\Omega)^2+\Gamma^2}\,,
\label{eq:fit}
\end{equation}
where $\Lambda\,,\Omega\,,\Gamma$ are fitting parameters representing a normalization constant, position of the maximum, and linewidth, respectively. As is evident from the presented results, for $q/\pi>0.5$, the magnons have a lifetime that is almost an order of magnitude smaller for $n\lesssim 1.5$ than for $n>1.5$. Surprisingly, for $q\to\pi$, some results (e.g., for $n=1.5$) regain coherence (at least partially).

\subsection{Magnon decoherence \& Stoner continuum}
\label{sec:magston}

The anomalous dependence of magnon lifetime on the electron density $n$ and wavelength $q$ indicates a nontrivial scattering of the magnons. The usual mechanism for such behavior in itinerant magnets emerges from the Stoner continuum. In this scenario, the magnetic excitations interact with charge fluctuations, i.e., scattering between two coherent bands of electrons below and above the Fermi level. In the generic case, one usually considers a polarized (or partially polarized) system and transitions between majority and minority electrons (which are parallel and antiparallel to the polarization of the system, respectively), modeled by, e.g., Eq.~\eqref{eq:ff} $\omega_\mathrm{ff}(q)$. However, these transitions correspond to $\omega\propto J_\mathrm{H}$, since majority and minority bands emerge from the mean-field decoupling of the Hund term in~\eqref{eq:hamkh}, i.e., $2J_\mathrm{H}\,\mathbf{s}_{\ell} \cdot \mathbf{S}_{\ell}\to2J_\mathrm{H}\,\left(s^z_{\ell} \cdot \langle S^z_{\ell}\rangle+\langle s^z_{\ell}\rangle \cdot S^z_{\ell}\right)$, justified only in $S\to\infty$ limit. Consequently, the Stoner continuum in such a scenario lies much above the energy span $W_\mathrm{S}\lesssim t/2$ of the spin excitations [see Fig.~\ref{fig:width}(a)]. Even without polarization, the charge fluctuations between states of $\omega_\mathrm{ff}(q)$ below and above $\epsilon_\mathrm{F}$ yield too high frequencies. In App.~\ref{app:charge}, we present such a situation which, for the free fermion system, corresponds to the dynamical charge structure factor $N(q,\omega)$. However, as discussed in Sec.~\ref{sec:charge}, the noninteracting (spinless) solution does not fully capture the electron dynamics; i.e., additional incoherent states exist below the Fermi level. The following will discuss how a Stoner-like picture emerges from this context.

\begin{figure}[!htb]
\includegraphics[width=1.0\columnwidth]{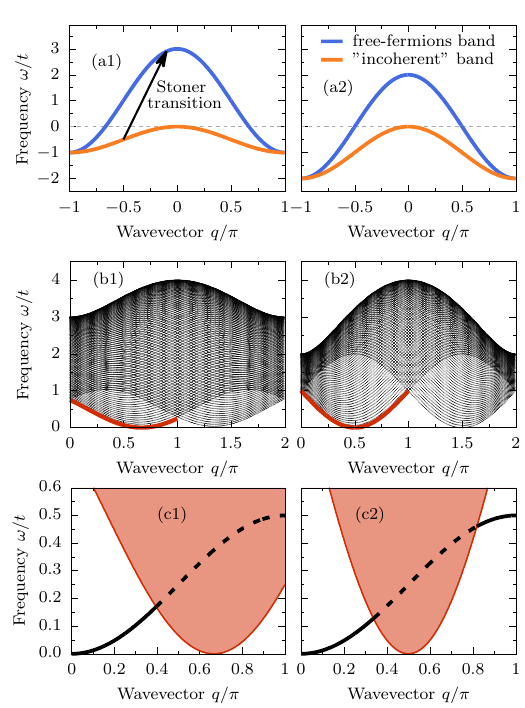}
\caption{Stoner continuum toy model (left column $n=1.33$, right column $n=1.50$). Panels (a1,a2) show a noninteracting band of coherent electrons $\omega_\mathrm{ff}(q)$, Eq.~\eqref{eq:ff}, and an approximation for the incoherent spectrum $\omega_\mathrm{inco}(q)$, Eq.~\eqref{eq:inco}. Panels (b1,b2) show the Stoner-like continuum $\omega_\mathrm{S}(q)$ relevant for transitions from coherent to incoherent bands, Eq.~\eqref{eq:stoner}. The red line depicts the minimum of the continuum $\omega_\mathrm{BS}(q)$, Eq.~\eqref{eq:stonerbootom}. (c1,c2) The low-frequency behavior of the Stoner continuum (solid color) and an exemplary coherent magnon dispersion $\omega_\mathrm{m}(q)=t/2[1-\cos(q)]$ (black line). The dashed line depicts the overlap of coherent magnon and Stoner continuum, i.e., $\omega_\mathrm{m}(q)>\omega_\mathrm{BS}(q)$.}
\label{fig:stoner}
\end{figure}

Building the Stoner-like continuum from an incoherent spectrum, like the one presented in Fig.~\ref{fig:akw_polarized}(b), requires some approximation. Let's consider a simple dispersion
\begin{equation}
\omega_\mathrm{inco}(q)=\omega_\mathrm{ff}(\pi)\big[1-\cos^2(q/2)\big]\,,
\label{eq:inco}
\end{equation}
from the bottom of the noninteracting band $\omega_\mathrm{inco}(\pi)=\omega_\mathrm{ff}(\pi)$ to the Fermi level $\omega_\mathrm{inco}(q=0)=0$. In the next paragraph, we will discuss the validity of this approximation. Here, let us first focus on the generic properties of the above toy model Eq.~\eqref{eq:inco}, also shown in Fig.~\ref{fig:stoner}(a1,a2) for two different densities $n$. The Stoner-like continuum can be constructed as
\begin{equation}
\omega_\mathrm{S}(q)=\omega_\mathrm{ff}(k_1)-\omega_\mathrm{inco}(k_2)\,,
\label{eq:stoner}
\end{equation}
where $q=\mod(k_1+k_2,2\pi)$, $k_1>k_\mathrm{F}$, and $k_2<k_\mathrm{F}$, see Fig.~\ref{fig:stoner}(b1,b2). Note that in our consideration [e.g., fully polarized $T^z_\mathrm{tot}=SL+sL(2-n)$ magnetization sector] $\omega_\mathrm{ff}(k_1)$ represents the band of $\sigma=\downarrow$ fermions, while $\omega_\mathrm{inco}(k_2)$ the incoherent band of $\sigma=\uparrow$ fermions. For Eq.~\eqref{eq:inco}, a simple analytical formula for the $q$-dependent minimum of the Stoner band from Eq.~\eqref{eq:stoner} is given by 
\begin{equation}
\omega_\mathrm{BS}(q)=\big(2t+\mu_\mathrm{ff}\big)\sin^2\big([q+k_\mathrm{F}]/2\big)\,.
\label{eq:stonerbootom}
\end{equation}
In Fig.~\ref{fig:stoner}(c1,c2), the low-frequency behavior of Eq.~\eqref{eq:stonerbootom} is contrasted with the expected dispersion of magnons, \hbox{$\omega_\mathrm{m}(q)\propto [1-\cos(q)]$}. The presented results show that the Stoner continuum does not affect the $\omega\to0$ physics, i.e., $\omega(q\to0)\propto q^2$. On the other hand, for $q\sim\pi/2$, the magnon dispersion overlaps with the Stoner continuum, and one expects {\it incoherent spin excitations in this region}, among the primary novel results of this publication. Interestingly, there are parameter regions where the $q\to\pi$ magnons can exit the Stoner continuum, regaining coherence [see Fig.~\ref{fig:stoner}(c2)].

\begin{figure}[!htb]
\includegraphics[width=1.0\columnwidth]{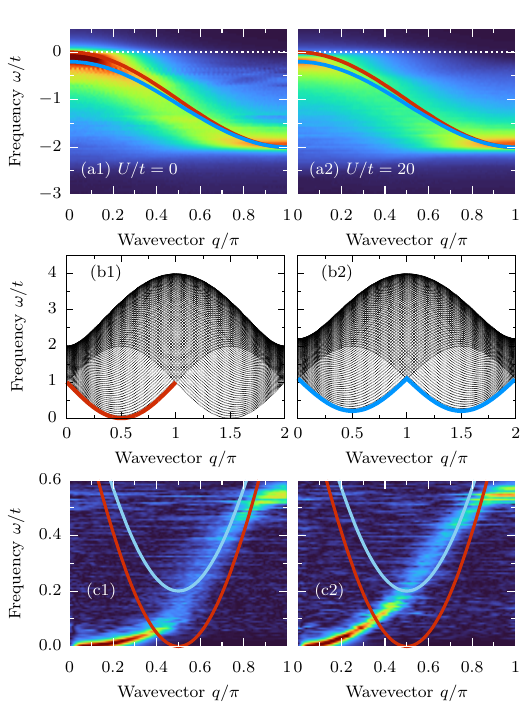}
\caption{(a1,a2) Incoherent part of the single-particle spectra calculated for $U/t=0$ (left column) and $U/t=20$ (right column). The red line represents the gapless approximation for the incoherent spectrum, Eq.~\eqref{eq:inco}, while the blue line represents the gapped one, i.e., Eq.~\eqref{eq:incogap} with $\Delta_\mathrm{inco}/t=0.1$. (b1,b2) The Stoner continuum obtained from the gapless (left column) and gapped (right column) $\omega_\mathrm{inco}(q)$. (c1,c2) Comparison of the $q$-dependent minimum of the Stoner continuum and the dynamical spin structure factor $S(q,\omega)$.}
\label{fig:stoner_gap}
\end{figure}

The behavior described above, see Fig.~\ref{fig:stoner}(c1,c2), is similar to the dynamical spin structure factor data presented in Fig.~\ref{fig:sqw}, particularly the nontrivial dependence of the magnon coherence on the electron density $n$. Before directly comparing with $S(q,\omega)$, let us first focus on the validity of Eq.~\eqref{eq:inco}. First, because the relevant energy scale of spin excitations is small, $W_\mathrm{S}\lesssim t/2$, only the limit $\omega_\mathrm{inco}\to 0$ is relevant for the spin dynamics.
Thus, almost the same results for $S(q,\omega)$ are obtained in the case when the incoherent excitations are approximated by the quadratic dispersion relations $\omega_\mathrm{inco}\propto -q^2$ (not shown). 

\begin{figure*}[!htb]
\includegraphics[width=1.0\textwidth]{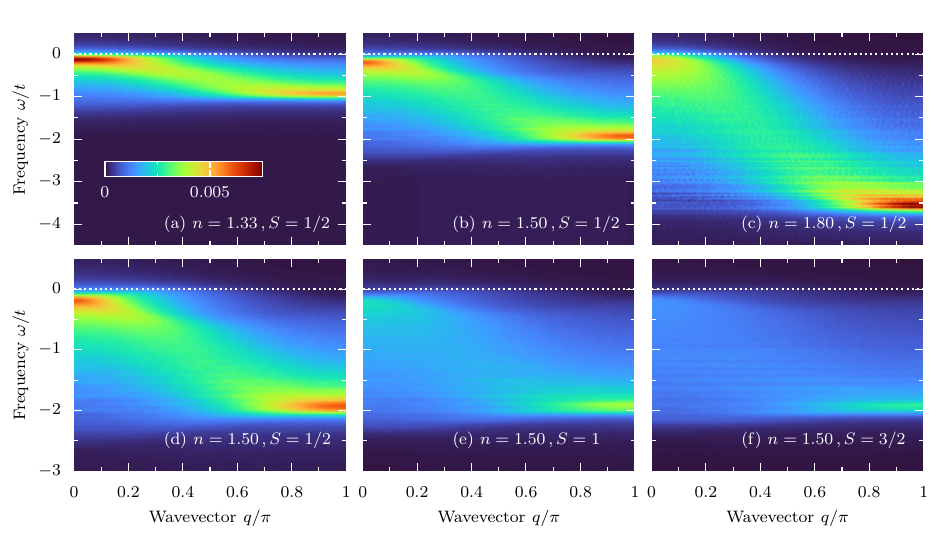}
\caption{Incoherent part of the single-particle spectral function $A(q,\omega)$, calculated for $\sigma=\uparrow$ electrons of the $\uparrow$-polarized system. Panels (a), (b), and (c) depict results for $S=1/2$ localized spins and $n=1.33\,,1.50\,,1.80$, respectively. Panels (d), (e), and (f) depict results for $n=1.50$ and $S=1/2\,,1\,,3/2$ localized spins, respectively. In all panels: $U/t=J_\mathrm{H}/t=20$, $T^z_\mathrm{tot}=SL+sL(2-n)$, $L=200$, $\delta\omega/t=4\cdot 10^{-2}$ and $\eta=2\delta\omega$.}
\label{fig:akw_inco}
\end{figure*}

Taking into account a particular feature of the incoherent states allows one to improve further the agreement between the toy model and the numerical results for the spin structure factor. Namely, our analysis indicates the presence of a small gap $\Delta_\mathrm{inco}$ in the incoherent part of the single-particle spectrum. In Fig.~\ref{fig:stoner_gap}(a1,a2) we show $J_\mathrm{H}\gg t$ results with $U/t=0$ and $U/t=20$. For the former, we observe that the incoherent $A(q,\omega)$ spectrum touches the Fermi level ($\Delta_\mathrm{inco}=0$), while for the latter we find $\Delta_\mathrm{inco}/t\simeq0.1$. Results presented in the App.~\ref{app:para} indicate a linear dependence on the Hubbard interaction strength, $\Delta_\mathrm{inco}\simeq \alpha U$, albeit with a very small coefficient $\alpha\ll 1$. The gap can be incorporated into the incoherent spectrum via the approximation
\begin{equation}
\omega_\mathrm{inco}(q)=\omega_\mathrm{ff}(\pi)\big[1-(1-\Delta_\mathrm{inco}/t)\cos^2(q/2)\big]\,,
\label{eq:incogap}
\end{equation}
yielding also an opening of a gap in the Stoner continuum, with $\min(\omega_\mathrm{BS})=2\Delta_\mathrm{inco}$; see Fig.~\ref{fig:stoner_gap}(b1,b2). A direct comparison of the spin spectrum $S(q,\omega)$ and the $q$-dependent minimum of the Stoner continuum calculated using Eq.~\eqref{eq:inco} and Eq.~\eqref{eq:incogap} is shown in Fig.~\ref{fig:stoner_gap}(c1,c2). A good qualitative agreement is observed for the gapless solution for both cases ($U/t=0$ and $U/t=20$). This holds particularly true for the \mbox{$U/W=0$} case (Fig.~\ref{fig:stoner_gap}, left column), where the bottom of the Stoner continuum derived from Eq.~\eqref{eq:inco} aligns perfectly with the incoherent part of the magnon spectrum. When considering finite interaction $U/t=20$, we observe better agreement with the gapped $\omega_\mathrm{inco}(q)$, achieving even quantitative agreement. 

In Fig.~\ref{fig:sqw}, we show a comparison of $S(q,\omega)$ and the Stoner continuum based on Eq.~\eqref{eq:incogap} with $\Delta_\mathrm{inco}/t=0.1$ (white dashed line). We find excellent agreement for all electron densities $n\leq 1.5$: (i) we observe coherent excitations at $q\to 0$. Next, (ii) for $\omega_\mathrm{m}(q)>\omega_\mathrm{BS}(q)$, i.e., in the region where magnons can interact with the Stoner continuum, we find that the former loses coherence even by one order of magnitude. Finally, (iii) for $n\sim1.5$, the dispersion partially regains coherence for short-wavelengths, $q\to\pi$.

However, the behavior of the spin excitation spectrum for $n>1.5$ is different. In this region, we find robust magnon mode softening for $q\to\pi$ and, consequently, a lack of any overlap with the Stoner continuum. This is consistent with the magnon linewidth $\Gamma$ result, which shows constant coherence for all wavevectors $q$; see Fig.~\ref{fig:width}(b). It is important to note that in this region, the approximation $\omega_\mathrm{inco}(q)$, Eq.~\eqref{eq:inco} or Eq.~\eqref{eq:incogap}, is no longer valid. In Fig.~\ref{fig:akw_inco}(a-c), we present the incoherent part of the single-particle spectrum for $n=1.33\,,1.50$, and $n=1.80$ (calculated from $\sigma=\uparrow$ electrons $\langle\langle c^{\dagger}_{\ell\uparrow} c^{\phantom{\dagger}}_{L/2\uparrow} \rangle\rangle_{\omega}^{+}$ of the $\uparrow$-polarized system). As already discussed, only the $\omega_\mathrm{inco}(q)<W_\mathrm{S}<t$ energy region is relevant for spin excitations (i.e., the bottom of the Stoner continuum). In the latter, $n \lesssim 1.5$ results have large spectral weight and can be approximated with $\omega_\mathrm{inco}(q)$. On the other hand, for $n>1.5$, we observe a continuum of excitations without any apparent structure close to the Fermi level. The naively assumed sharp dispersion, $\omega_\mathrm{inco}(q)$, from the simplistic Stoner analysis, is not observed in this regime when all degrees of freedom are accurately considered. Consequently, in order to properly evaluate the magnon dispersion relation in this regime, one needs to consider a more fundamental Hamiltonian than in previous investigations, as the one in our effort.

\begin{figure*}[!htb]
\includegraphics[width=1.0\textwidth]{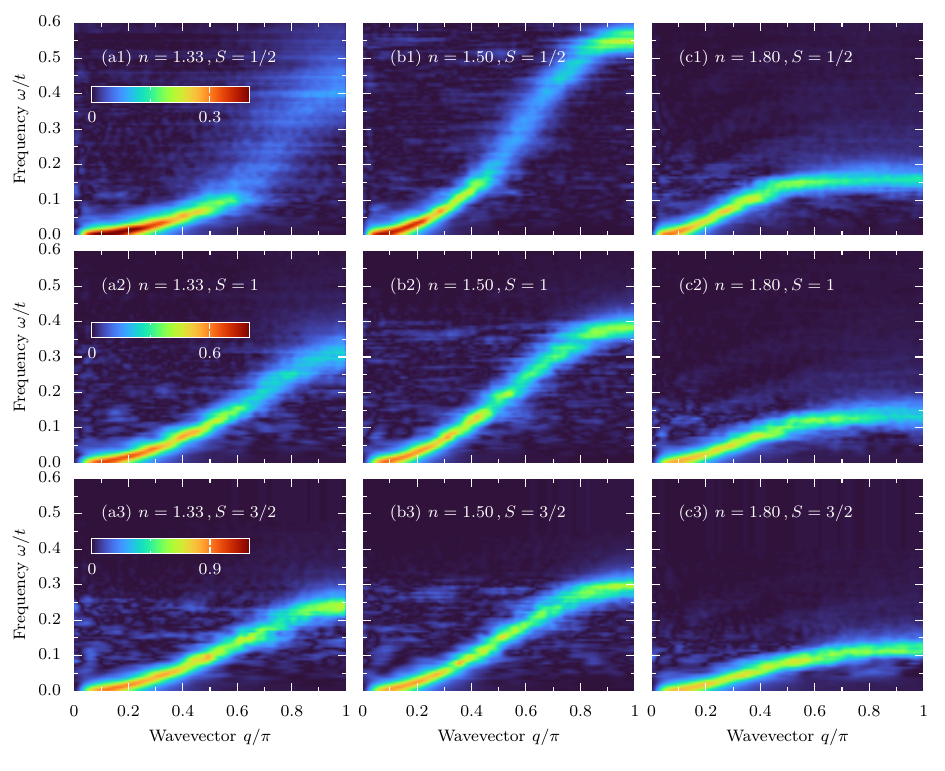}
\caption{Dynamical spin structure factor $S(q,\omega)$ calculated for various magnitudes of localized spins $S$ and various electronic densities $n$ in the limit of $J_\mathrm{H}\gg t$ ($U/t=J_\mathrm{H}/t=20$ and $T^z_\mathrm{tot}=0$ magnetization sector). (a1-a3) $n=1.33$, (b1-b3) $n=1.50$, and (c1-c3) $n=1.80$. The first, second, and third rows represent results for $S=1/2$, $S=1$, and $S=3/2$, respectively. In all panels: $L=200$, $\delta\omega/t=6\cdot 10^{-3}$ and $\eta=2\delta\omega$.}
\label{fig:sqw_S}
\end{figure*}

\subsubsection*{Spin magnitude dependence}
\label{sec:spinlen}

So far, we have considered the gK model~\eqref{eq:hamkh} with $S=1/2$ localized spins. Such considerations are relevant for, e.g., iron-based materials in the OSMP phase, where one of the orbitals is Mott localized. Conversely, manganites are typically described with $S=3/2$, arising from three localized electrons in $t_{2g}$ orbitals. In this Section, we will investigate the dependence of spin excitations $S(q,\omega)$ on the magnitude of the localized spin $S$. In Fig.~\ref{fig:sqw_S}, we present the magnon spectrum for various electron densities, $n=1.33\,,1.50\,,1.80$, and $S=1/2\,,1\,,3/2$. Our results indicate that for large densities, $n=1.80$, the magnitude of the localized spin $S$ has little effect on the overall magnon dispersion $\omega_\mathrm{m}(q)$.

On the other hand, two important effects of the magnitude of localized $S$ on spin excitations occur for $n \lesssim 1.5$. Firstly, the magnon dispersion softens for short wavelengths ($q \to \pi$) for all $n$, a result similar to the $S=1/2, n=1.80$ finding. Secondly, as $S$ increases, we observe that the magnons regain coherence for all wavevector values $q$. The analysis of the single-particle spectra $A(q,\omega)$, shown in Fig.~\ref{fig:akw_inco}(d-f), indicates that the incoherent part does not have a well-defined structure for $\omega\to0$. Consequently, $\omega_\mathrm{inco}(q)$, Eq.~\eqref{eq:inco}, is not a good approximation since there are no well-defined states from which the Stoner continuum can be built, again similar to the $S=1/2$ case with $n=1.80$. In addition, in panels (d)-(f) of Fig.~\ref{fig:akw_inco}, we see how the weight of the incoherent band in $A(q,\omega)$ decreases as $S$ increases, indicating that as $S\to\infty$ (limit of classical spins) this weight disappears~\cite{Yunoki1998}.

\subsection{Magnon mode softening}
\label{sec:magsoft}

\begin{figure}[!htb]
\includegraphics[width=1.0\columnwidth]{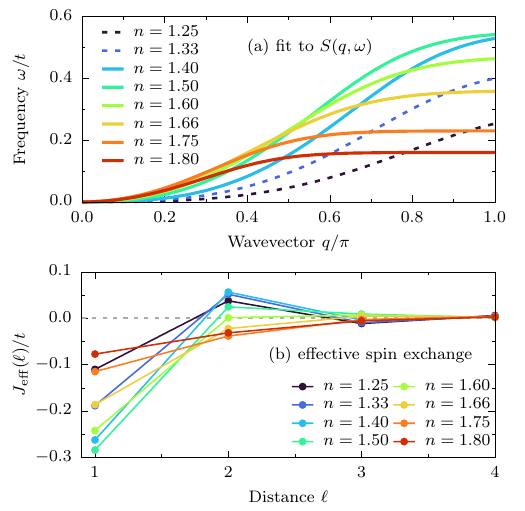}
\caption{(a) Magnon dispersion $\omega_\mathrm{fit}(q)$ given by the fit to $S(q,\omega)$ for various electron densities $n$ (see also black dashed lines in Fig.~\ref{fig:sqw}). The dashed lines for $n<1.4$ indicate the region where the fit is less accurate due to strong magnon decoherence. (b) Spin exchange $J_\mathrm{eff}(\ell)$ of the effective Heisenberg model~\eqref{eq:heis} obtained from Fourier transform of the dispersion presented in panel (a).}
\label{fig:longrangeJ}
\end{figure}

The Stoner continuum $\omega_\mathrm{S}(q)$ described in the previous section does not explain the magnon mode softening for $q\to\pi$ observed for $n>1.5$. In fact, our results indicate very small decoherence in this region; see Fig.~\ref{fig:width}(b). This is consistent with the overall analysis presented in Sec.~\ref{sec:magston}: (i) our results presented in Fig.~\ref{fig:sqw}(d-f) indicated that the Stoner continuum lies above the spin excitations due to the presence of the $\Delta_\mathrm{inco}$ gap. (ii) Even if we consider $\Delta_\mathrm{inco}=0$ (i.e., $U=0$ case), in $\omega\to0$ limit there is no well-defined spectral weight of the incoherent part of $A(k,\omega)$, see Fig.~\ref{fig:akw_inco}(c), and $\omega_\mathrm{inco}(q)$ is not a good approximation, as discussed earlier. 

In this Section, we will discuss the spin exchange interaction in the effective Heisenberg-type model 
\begin{equation}
H_\mathrm{eff}=\sum_{i,\ell}J_\mathrm{eff}(\ell)\,\mathbf{S}_{i}\cdot\mathbf{S}_{i+\ell}\,,
\label{eq:heis}
\end{equation}
with which the dynamical spin structure factor $S(q,\omega)$ can be described, especially the magnon mode softening phenomenon. Note that the spin length $S$ in the above model is not crucial for ferromagnetically ordered systems, i.e., the results do not change when rescaled exchange $J_\mathrm{eff}\to J_\mathrm{eff}/S$ is considered (as evident from, e.g., the general form of Holstein–Primakoff transformation yielding the standard $JS[1-\cos(q)]$ dispersion). Furthermore, although one does not expect the decoherence of magnons in $H_\mathrm{eff}$, the general shape of the dispersion $\omega_\mathrm{S}(q)$ can be captured by a proper choice of $J_\mathrm{eff}(\ell)$ - the approach which is an essence of the spin-wave theory considerations.

\begin{table}[!htb]
\centering 
\begin{tabular}{|c||c|c|c|}
\hline
$\quad n\quad$ & $\quad J_\mathrm{eff}(1)/S/t\quad$ & $\quad J_\mathrm{eff}(2)/S/t\quad$ & $\, J_\mathrm{eff}(2)/J_\mathrm{eff}(1)\,$ \\ \hline\hline
$1.25$ & $-\,0.055$ & $+\,0.018$ & $-\,0.343$ \\ \hline
$1.33$ & $-\,0.094$ & $+\,0.026$ & $-\,0.275$ \\ \hline
$1.40$ & $-\,0.131$ & $+\,0.028$ & $-\,0.216$ \\ \hline
$1.50$ & $-\,0.142$ & $+\,0.012$ & $-\,0.087$ \\ \hline
$1.60$ & $-\,0.121$ & $+\,0.001$ & $-\,0.001$ \\ \hline
$1.66$ & $-\,0.093$ & $-\,0.011$ & $+\,0.119$ \\ \hline
$1.75$ & $-\,0.057$ & $-\,0.019$ & $+\,0.333$ \\ \hline
$1.80$ & $-\,0.039$ & $-\,0.016$ & $+\,0.407$ \\ \hline
\end{tabular}
\caption{Nearest- and next-nearest-neighbor spin exchange, $J_\mathrm{eff}(\ell=1,2)$, of the effective Heisenberg model~\eqref{eq:heis}. The last column represents the absolute value of the $J_\mathrm{eff}(2)/J_\mathrm{eff}(1)$ ratio.}
\label{tab:j1j2}
\end{table}

In order to estimate the $J_\mathrm{eff}(\ell)$, we fit the collection of frequencies for which the $S(q,\omega)$ takes maximum value for given wavevector $q$. We find the most consistent fit for all electron densities $n$ can be obtained with \mbox{$\omega_\mathrm{fit}(q)=a\tanh(b\,q^c)$}, with $a,b,c$ as fit parameters. Note that the functional form of $\omega_\mathrm{fit}(q)$ is here arbitrary, i.e., our aim is to mimic the $q$-dependence of the $S(q,\omega)$ maximum in the whole range of wavevectors $0<q<\pi$, even in the strongly incoherent region. The result of such a procedure is presented as a black dashed line in Fig.~\ref{fig:sqw}, while the details are presented in the App.~\ref{app:fit}. Note that the $\omega_\mathrm{fit}(q)$ is a crude approximation for $n<1.4$ data since the strong decoherence of excitations for $q/\pi>0.5$ prevents an accurate fit. Nevertheless, our results indicate a systematic variation with $n$ even in this region [see the summary of the results for various $n$ presented in Fig.~\ref{fig:longrangeJ}(a)]. 

\begin{figure}[!htb]
\includegraphics[width=1.0\columnwidth]{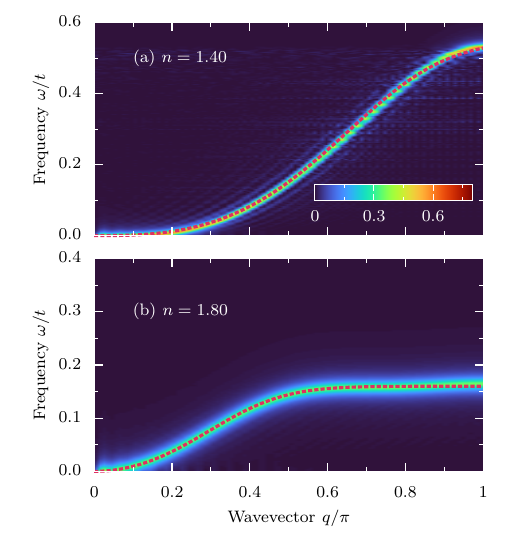}
\caption{Dynamical spin structure factor $S(q,\omega)$ of the effective Heisenberg model as calculated for $S=1/2$ and values of spin exchange $J_\mathrm{eff}(\ell=1,2)$ given by Tab.~\ref{tab:j1j2} for (a) $n=1.40$ and (b) $n=1.80$ (and $J_{\mathrm{eff}}(\ell\geq3)=0$ otherwise). The red dashed lines represent the fitted $\omega_\mathrm{fit}(q)$ dispersion for given $n$. In all panels: $L=100$, $\delta\omega/t=6\cdot 10^{-3}$ and $\eta=2\delta\omega$.}
\label{fig:heisenberg}
\end{figure}

The estimate for $J_\mathrm{eff}(\ell)$ is then obtained from Fourier transform of $\omega_\mathrm{fit}(q)$. Our results presented in Fig.~\ref{fig:longrangeJ}(b) indicate that (i) nearest-neighbor (NN) exchange is negative $J_\mathrm{eff}(1)<0$ for all $n$, as expected for ferromagnetically ordered systems. (ii) Furthermore, effective spin exchanges decay fast beyond next-NN, $J_\mathrm{eff}(\ell\geq3)\simeq 0$. Consequently, the fitted dispersion relation can be reproduced just with NN and next-NN interaction (see Tab.~\ref{tab:j1j2}). In Fig.~\ref{fig:heisenberg}, we show exemplary dynamical spin structure factor $S(q,\omega)$ of the effective Heisenberg model~\eqref{eq:heis} with $S=1/2$ and values of $J_\mathrm{eff}$ given in Tab.~\ref{tab:j1j2} and $J_\mathrm{eff}(l\geq3)=0$. As evident from the presented results for $n=1.4$, the maximum of $S(q,\omega)$ accurately follows $\omega_\mathrm{fit}(q)$. However, the original spin structure factor presented in Fig.~\ref{fig:sqw} is reproduced only qualitatively since there is no magnon decoherence in the Heisenberg model~\eqref{eq:heis}. On the other hand, for $n=1.8$, we obtain a quantitative agreement not only with $\omega_\mathrm{fit}(q)$ but also with the magnons in the full model $H_\mathrm{gK}$,~\eqref{eq:hamkh}.

Interestingly, our $J_\mathrm{eff}$ results indicate the change in the nature (sign) of $J_\mathrm{eff}(\ell=2)$ with density $n$. For $n\lesssim 1.6$ the sign of spin exchange is AFM ($J_\mathrm{eff}(2)>0$), while for $n\gtrsim 1.6$ FM ($J_\mathrm{eff}(2)<0$). This behavior coincides with the change in the slope of $J_\mathrm{q\to0}$ presented in Fig.~\ref{fig:width} and, more importantly, in the change in the behavior of double-exchange magnons. Our analysis in the previous section indicates that for $n\lesssim1.6$, the magnons strongly scatter on the Stoner continuum of incoherent electrons, while for $n\gtrsim 1.6$, one observes the magnon mode softening. Finally, it is worth noting that the ratio $|J_\mathrm{eff}(2)/J_\mathrm{eff}(1)|$ takes large values for extremes $n$, i.e., for small $n=1.25$ and large $n=1.80$ doping. Such ratios are consistent with experimental estimates on spin exchanges \cite{Zhang2007} (note that the fourth-neighbor exchange used in the three-dimensional model corresponds to $J_\mathrm{eff}(2)$ for the 1D lattice dimensionality).

As a final remark of this section, we want to comment on the relation between $J_\mathrm{eff}$ and the electron correlations. The spin-exchange interactions in effective spin models derived from Kondo lattice-like Hamiltonians are typically linked to the kinetic energy of the conduction electrons (reminiscent of the second scenario for magnetism discussed in the introduction \cite{Moriya1979}). For example, for $J_\mathrm{H}\to0$, the long-range spin exchange mediated by itinerant electrons - the so-called Ruderman–Kittel–Kasuya–Yosida (RKKY) interaction - can be perturbatively derived \cite{Ruderman1954,Kasuya1956,Yosida1957}. In the limit of $S\to\infty$, the $1/S$ spin-wave expansion \cite{Shannon2002,Furukawa2004} relates $J_\mathrm{eff}(1)$ to the average kinetic energy per bond and $t/J_\mathrm{H}$ corrections (in $J_\mathrm{H}/t\gg 1$ limit) can also induce $J_\mathrm{eff}(2)$ coupling \cite{Frakulla2024}. However, in the latter case, $J_\mathrm{eff}(2)\propto1/J_\mathrm{H}$, i.e., are small and of the same sign as $J_\mathrm{eff}(1)$. Such considerations can't capture the phenomena shown in Fig.~\ref{fig:longrangeJ} and Tab.~\ref{tab:j1j2}.

\begin{figure}[!htb]
\includegraphics[width=1.0\columnwidth]{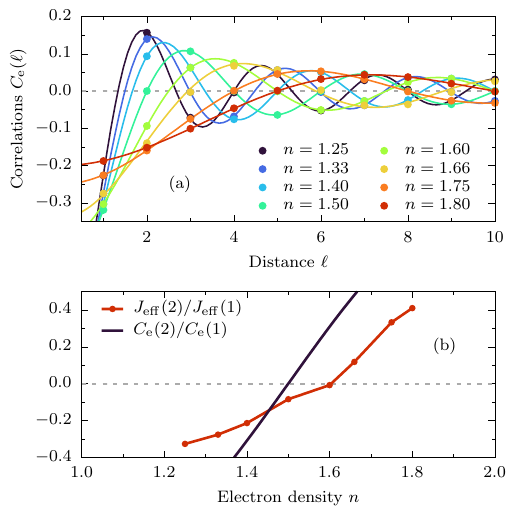}
\caption{(a) Static electron correlations $C_\mathrm{e}(\ell)\propto \langle c^\dagger_{i} c^{\phantom{\dagger}}_{i+\ell}\rangle$ of gK model with $S=1/2$ localized spins, calculated for \mbox{$U/t=J_\mathrm{H}/t=20$}, $T^z_\mathrm{tot}=0$, and $n=1.25\,,\dots,1.80$ (points). Lines represent noninteracting spinless fermions solution $C(\ell)=-\sin(k^\mathrm{ff}_\mathrm{F}\ell)/\pi/\ell$ with $k^\mathrm{ff}_\mathrm{F}=\pi n_\mathrm{ff}=\pi(2-n)$. (b) Electron density $n$ dependence of the ratio between nearest- and next-nearest neighbor effective spin exchanges $J_\mathrm{eff}(2)/J_\mathrm{eff}(1)$ and electron correlations $C_\mathrm{e}(2)/C_\mathrm{e}(1)=\cos(k^\mathrm{ff}_\mathrm{F})$.}
\label{fig:cdc}
\end{figure}

In electron-mediated spin exchange scenarios, $J_\mathrm{eff}(\ell)$ should be proportional to the static electron correlations \cite{Furukawa2004} of the form
\begin{equation}
C_\mathrm{e}(\ell)=\frac{1}{L-\ell} \sum_{\sigma,i=1}^{L-\ell} 
\langle c^\dagger_{i\sigma} c^{\phantom{\dagger}}_{i+\ell\sigma}\rangle\,.
\end{equation}
Note that $C_\mathrm{e}(1)$ is proportional to the kinetic energy of the system. Consistently with the previous discussion, the results presented in Fig.~\ref{fig:cdc}(a) yield that the behavior of electrons is that of noninteracting spinless fermions at $n_\mathrm{ff}=2-n$ density. For the latter, the electron correlations are given by $C_\mathrm{e}(\ell)=-\sin(k^\mathrm{ff}_\mathrm{F}\ell)/\pi/\ell$ with $k^\mathrm{ff}_\mathrm{F}=\pi n_\mathrm{ff}$ (the result with which we are in perfect agreement). Within such a solution, the ratio between nearest- and next-nearest neighbor correlations is given by $C_\mathrm{e}(2)/C_\mathrm{e}(1)=\cos(n^\mathrm{ff}_\mathrm{F})$. In Fig.~\ref{fig:cdc}(b) we contrast the latter and $J_\mathrm{eff}(2)/J_\mathrm{eff}(1)$ value obtained from the fits to the dispersion. As evident, $C_\mathrm{e}$ only qualitatively captures the effective spin exchange behavior, i.e., it captures the overall change of sign of $J_\mathrm{eff}(2)$ with $n$. However, the $C_\mathrm{e}(2)/C_\mathrm{e}(1)$ changes sign for $n=1.5$, while our data indicate the change in $J_\mathrm{eff}(2)/J_\mathrm{eff}(1)$ for $n\simeq 1.6$. Also, the electron density $n$ dependence of $C_\mathrm{e}(2)/C_\mathrm{e}(1)$ is much stronger than the one obtained from the fits. 

Our results validate the experimental observation \cite{Hwang1998,Endoh2005,Ye2006,Zhang2007} that describing magnon mode softening requires incorporating second-NN interactions along a primary lattice direction in the effective spin model. In the three-dimensional classical spin-wave consideration, this indicates strong spatial exchange anisotropy, i.e., finite coupling in $[1,0,0]\,,[0,1,0]\,,[0,0,1]$ and $[2,0,0]\,,[0,2,0]\,,[0,0,2]$ direction and vanishing in $[1,1,0]\,,[1,0,1]$ and $[1,1,1]$ direction. Various scenarios were proposed for the origin of this non-monotonic behavior, e.g., coupling to phonons \cite{Woods2001}, orbital ordering \cite{Endoh2005}, or breakdown of the canonical double-exchange limit \cite{Solovyev1999}. Here, we show that calculations within a fully quantum model reproduce the experimental findings.

\section{Two-orbital Hubbard model}
\label{sec:spintwoorb}

The generalized Kondo model~\eqref{eq:hamkh} with \mbox{$S=1/2$} localized moments is an effective description of the OSMP of the two-orbital Hubbard model \cite{Herbrych2019}. However, ferromagnetically ordered phases occur over a wide range of parameters in~\eqref{eq:ham2o}, even in the absence of apparent localized electrons \cite{Momoi1998,Sakai2001,Sakai2007,Kubo2009,Peters2010,LingFang2023}, that is, with finite charge fluctuations in all orbitals. In such situations, the spin excitation analysis must be performed in the full multiorbital setup.

\begin{figure}[!htb]
\includegraphics[width=1.0\columnwidth]{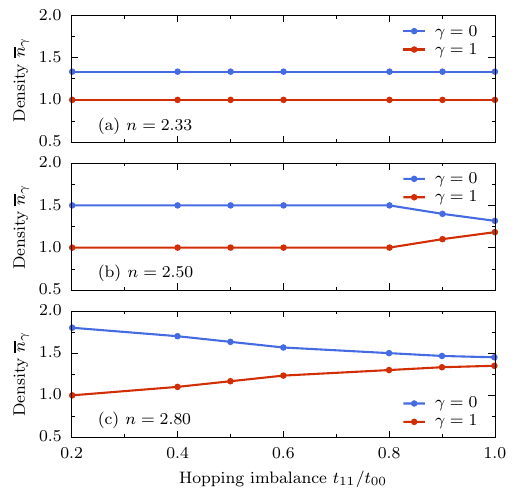}
\caption{Orbitally-resolved electron density $n_\gamma$ for $\gamma=0\,,1$ of the two-orbital Hubbard-Kanamori model as a function of orbital differentiation $t_{11}/t_{00}$. Calculated for $L=60$ sites, $U/t=32$, $J_\mathrm{H}=U/4$, and (a,b,c) $n=2.33\,,2.50\,,2.80$, respectively.}
\label{fig:hubbard_density}
\end{figure}

In this Section, we will consider the two-orbital Hubbard-Kanamori model~\eqref{eq:ham2o} with hopping $t_{00}=0.5$, $t_{01}=t_{10}=0$, varying $t_{11} \le t_{00}$, and crystal field $\Delta_\mathrm{CF}/t=0.2$. We expect that orbital differentiation, predominantly induced by $t_{11}\ne t_{00}$, leads to the OSMP phase for sufficiently large Hubbard and Hund interactions, as shown in previous studies \cite{Rincon2014,Herbrych2018,Herbrych2020, Ko2023}. Here, we choose a representative large Hubbard interaction $U/t=32$. In multiorbital systems, both the Hubbard and Hund values originate from Coulomb interactions \cite{Oles1983}. Consequently, we link these two parameters by the relation $J_\mathrm{H}=U/4$ \cite{Luo2010,Yin2011,Dai2012}. Finally, to match the results of this Section to the previous ones, we select a total electron density of $n=2.33\,,2.50\,,2.80$. If the system enters the OSMP, such $n$ will correspond to one electron in the Mott localized orbital ($\gamma=1$) and an electron density of $1.33\,,1.50\,,1.80$ on the itinerant orbital ($\gamma=0$), respectively. In Fig.~\ref{fig:hubbard_density} we present the hopping imbalance $t_{11}/t_{00}$ dependence of the orbital $\gamma=0,1$ resolved electron density $n_\gamma=\sum_\ell n_{\gamma\ell}$. Our results indicate that for the large enough orbital differentiation (here $t_{11}/t_{00}=0.2$), the $\gamma=1$ orbital is singly occupied for all considered values of $n$, indicating OSMP. In the opposite limit of equal bandwidth, $t_{00}=t_{11}$, for $n=2.50$ and $n=2.80$, both orbitals are fractionally occupied; see Fig.~\ref{fig:hubbard_density}(b,c). For $n=2.33$, the system is in the OSMP even for $t_{00}=t_{11}$ due to a finite crystal field-splitting $\Delta_\mathrm{CF}\ne0$.

\begin{figure*}[!htb]
\includegraphics[width=1.0\textwidth]{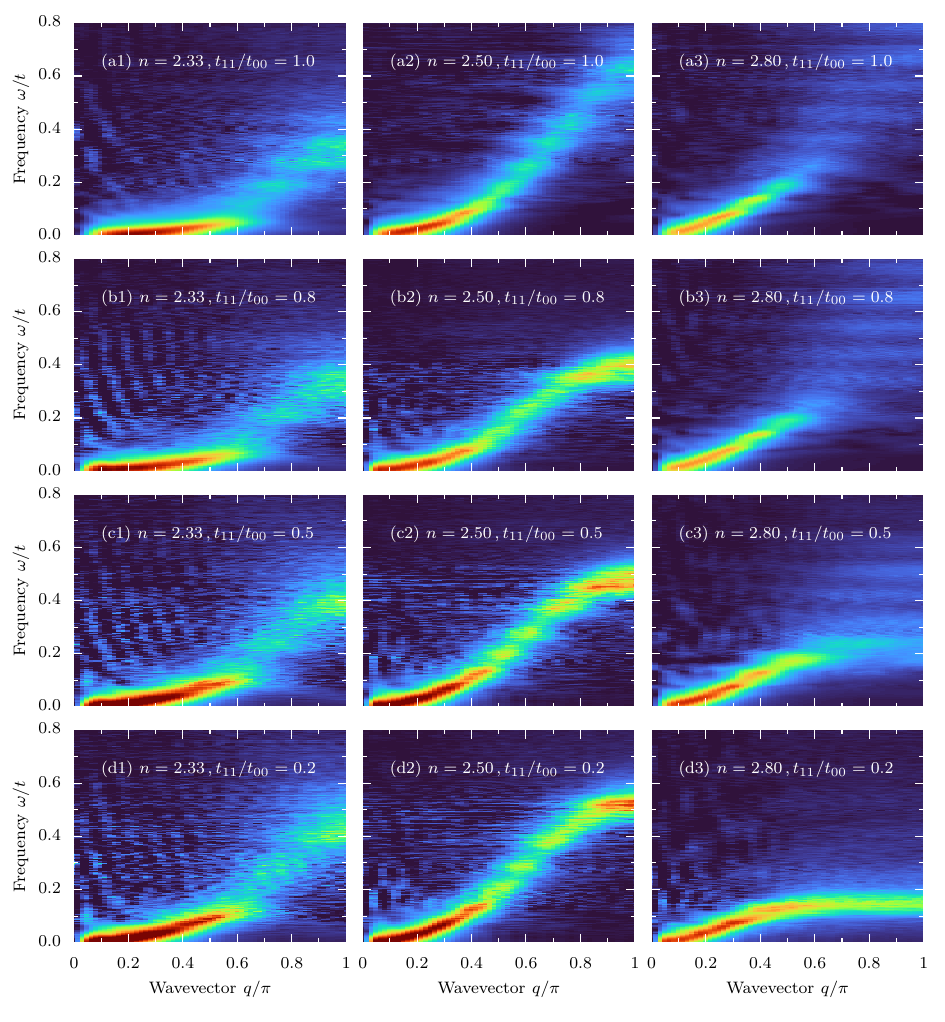}
\caption{Dynamical spin structure factor $S(q,\omega)$ of the two-orbital Hubbard-Kanamori (HK) model in the $U\,,J_\mathrm{H}\gg t$ limit for various hopping imbalances hopping imbalances $t_{11}/t_{00}=1.0\,,0.8\,,0.5\,,0.2$ (rows) and electronic densities $n=2.33\,,2.50\,,2.80$ (columns). In all panels: $L=60$, $t_{00}=0.5$, $\Delta_\mathrm{CF}/t=0.2$, $U/t=32$, $J_\mathrm{H}/U=1/4$, $\delta\omega/t=4\cdot 10^{-3}$ and $\eta=2\delta\omega$.}
\label{fig:sqw_hubbard}
\end{figure*}

The main result of this Section, namely the dynamical spin structure factor $S(q,\omega)$ of the two-orbital Hubbard model in the $U\,,J_\mathrm{H}\gg t$ limit, is shown in Fig.~\ref{fig:sqw_hubbard}. Here, we use $\mathbf{T}_{\ell}=\mathbf{S}_{0\ell}+\mathbf{S}_{1\ell}$ in Eq.~\eqref{eq:sqw}. For almost all considered parameters, we observe strong magnon decoherence. The exceptions are the results for $n=2.80$ and $t_{11}/t_{00}\gtrsim 0.5$, where magnon mode softening is visible. Note that the results for $n=2.33=1+1.33$ of the HK model~\eqref{eq:ham2o} are consistent with the results for $n=1.33$ of gK~\eqref{eq:hamkh} for all values of $t_{11}/t_{00}$ considered. On the other hand, for $n=2.80=1+1.80$, only the $t_{11}/t_{00}=0.2$ (i.e., in the OSMP region) dispersion is akin to the $n=1.80$ gK results.

\begin{figure*}[!htb]
\includegraphics[width=1.0\textwidth]{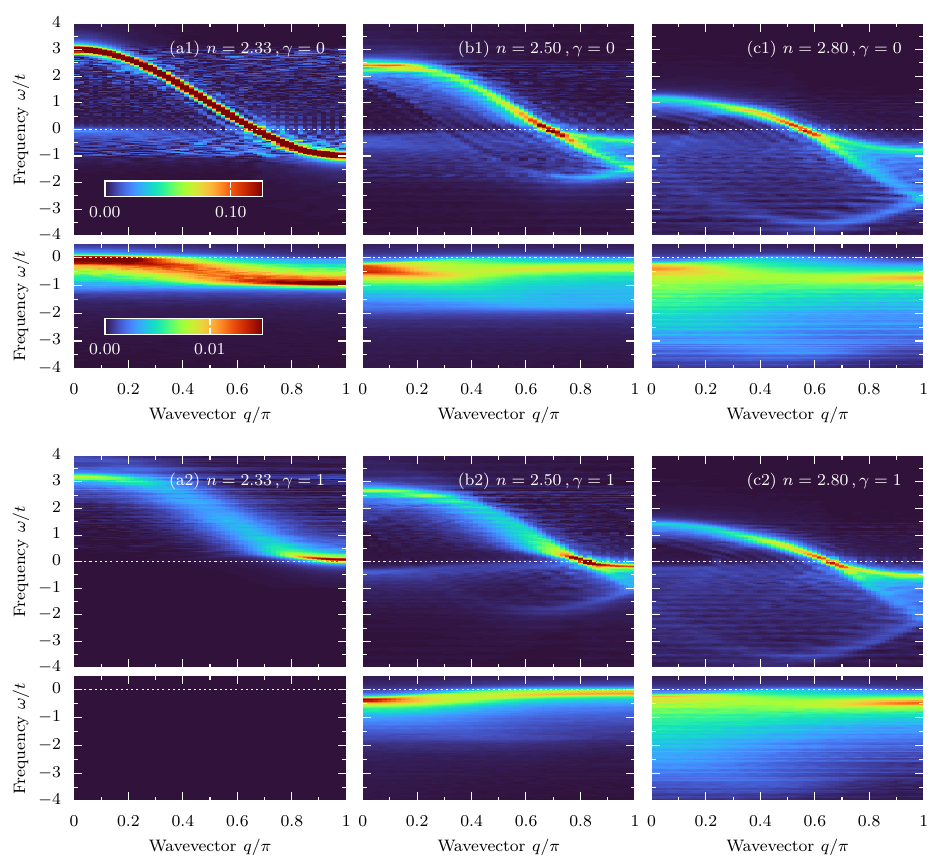}
\caption{Single-particle spectral function $A(q,\omega)$ in the $T^z_\mathrm{tot}=0$ sector (large panels) and the incoherent spectrum evaluated in the $\uparrow$-polarized system [$T^z_\mathrm{tot}=(4-n)L/2$] (small panels) calculated for the two-orbital HK model. The upper (lower) row represents the $\gamma=0$ ($\gamma=1$) orbital data. Panels (a,b,c) depict data for $n=2.33\,,2.50\,,2.80$, respectively. In all panels: $L=60$, $t_{11}/t_{00}=0.5$, $\Delta_\mathrm{CF}/t=0.2$, $U/t=32$, $J_\mathrm{H}/U=1/4$, $\delta\omega/t=4\cdot 10^{-2}$ and $\eta=2\delta\omega$.}
\label{fig:akw_hubbard}
\end{figure*}

This behavior can be easily understood through insights from the magnon dispersion analysis of the gK model presented in previous Sections. There, we observed a reduced magnon lifetime when the density of (itinerant) electrons was smaller than $1.5$, i.e., when the incoherent part of the spectrum had sharp features close to the Fermi level. We repeat this analysis for the HK model~\eqref{eq:hamkh} evaluating the single-particle spectral function $A(q,\omega)$ in the $T^z_\mathrm{tot}=0$ magnetization sector, as well as $A_{\uparrow}(q,\omega)$ in the $\uparrow$-polarized system [$T^z_\mathrm{tot}=(4-n)L/2$] to resolve the incoherent part of the spectrum. The results for $n=2.33\,,2.50\,,2.80$ and $t_{11}/t_{00}=0.5$ are shown in Fig.~\ref{fig:akw_hubbard}. Before we discuss our findings, two general remarks are necessary. Firstly, within the HK model, the single-particle spectra can be orbitally resolved [i.e., $c^{\phantom{\dagger}}_{\ell}\to c^{\phantom{\dagger}}_{\gamma\ell}$ and $c^{\dagger}_{\ell}\to c^{\dagger}_{\gamma\ell}$ in Eq.~\eqref{eq:akw}]. Secondly, it is evident that the spectral function $A(q,\omega)$ of HK model~\eqref{eq:hamkh} is far more complicated than the one presented in Fig.~\ref{fig:akw} for the gK model. For the cases beyond the OSMP regime, namely $n=2.50\,,2.80$, we do not observe any signature of well-defined quasiparticles [a $\delta(\omega)$-like spectral feature as in the case of Fig.~\ref{fig:akw}], at least within the limitations of our computational techniques. Instead, a broad spectrum known from generic strongly-correlated systems is found. To avoid confusion, in the following, we reserve the term incoherent spectrum for the part found in the $\uparrow$-polarized system by investigating the electrons with spin parallel to polarization $A_{\uparrow}(k,\omega)$, following the analysis presented in Sec.~\ref{sec:charge}. 

Consider first the data for $n=2.33$ and $t_{11}/t_{00}=0.5$ shown in the left column of Fig.~\ref{fig:akw_hubbard}. The $A(q,\omega)$ of the itinerant orbital $\gamma=0$ (also the incoherent part) is identical to the gK result at the corresponding filling ($n=1.33$); compare Fig.~\ref{fig:akw_hubbard}(a1) with Fig.~\ref{fig:akw}(a) and Fig.~\ref{fig:akw_inco}(a). In the localized orbital $\gamma=1$, Fig.~\ref{fig:akw_hubbard}(a2), the gap at the Fermi level opens, and no incoherent part of the spectrum is present, consistent with the OSMP prediction. The spin structure factor for such a density (shown in the left column of Fig.~\ref{fig:sqw_hubbard}) also agrees with the previous gK result, i.e., an incoherent magnon spectra develops for $q/\pi>0.6$. Since the $n=2.33$ results are always in the OSMP region for all considered parameters, the above behavior holds for all the $t_{11}/t_{00}$ cases.

The single-particle spectrum outside the OSMP, $n=2.50\,,2.80$ for $t_{11}/t_{00}=0.5$, is different. Here, both orbitals contribute to the states at the Fermi level $\omega=0$. For equal bandwidth, the occupation in both orbitals is approximately (due to the presence of small $\Delta_\mathrm{CF}\ne0$) equal to $n_\gamma\simeq n/2<1.5$, leading to the metallic nature of both orbitals. The gK analysis indicates that such densities should lead to the presence of the incoherent $A_{\uparrow}(q,\omega)$. The results presented in Fig.~\ref{fig:akw_hubbard}(b,c) indicate that, indeed, both orbitals contain an incoherent part below $\omega=0$ leading to incoherent magnon spectra in $q\to\pi$ limit; see Fig.~\ref{fig:sqw_hubbard}(a2,a3). 

\begin{figure}[!htb]
\includegraphics[width=1.0\columnwidth]{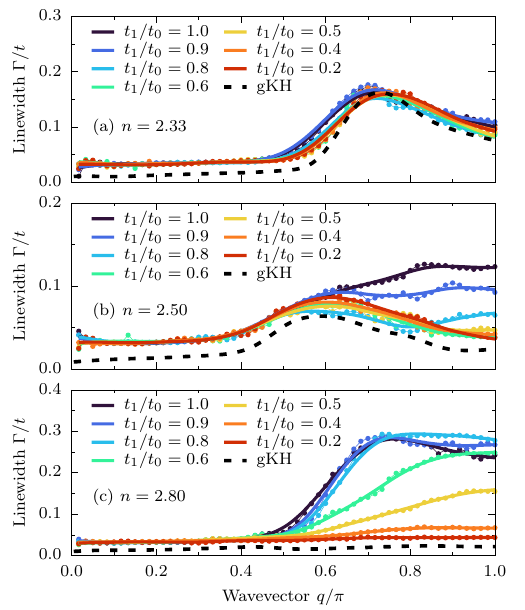}
\caption{Wavevector $q$ dependence of the magnon linewidth $\Gamma$ for various hopping imbalances $t_{11}/t_{00}$ of the HK model. Panels (a,b,c) represent the $n=2.33\,,2.50\,,2.80$ electron densities, respectively. Points represent the result of the fit, Eq.~\eqref{eq:fit}, while solid lines represent guides to the eye. The black dashed line represents the corresponding gK result; see Fig.~\ref{fig:width}(b).}
\label{fig:width_hubbard}
\end{figure}

Increasing the hopping imbalance, i.e., decreasing the ratio $t_{11}/t_{00}<1$, moves the system into the OSMP region when the electron density of the localized orbital $\gamma=1$ reaches unity, transferring the weight to the itinerant orbital $\gamma=0$. Our gK results indicate that the density on the latter orbital controls the incoherent spectra, which in turn influences the magnon behavior. In the case of $n=2.50$, this leads (within OSMP) to incoherent magnons within a finite range of $q$, similar to the $n=1.50$ results of gK. For $n=2.80$, magnon mode softening is observed, known from $n=1.80$ gK considerations. One can monitor this scenario by examining the behavior of the magnon linewidth $\Gamma$ depicted in Fig.~\ref{fig:width_hubbard}. For the $n=2.33$ case, we do not observe any significant change in $\Gamma$ with $t_{11}/t_{00}$ since the system remains in the OSMP. For other considered values of electron densities, initially ($t_{00}\simeq t_{11}$), a strong reduction of magnon lifetime is present. With increasing imbalance, the magnon regains partial coherence for $q\to\pi$ for $n=2.50$ or full coherence (and mode softening) for $n=2.80$. The results are consistent with the gK prediction in all cases, provided that the system is in OSMP (see dashed line in Fig.~\ref{fig:width_hubbard}).

\section{Discussion \& Conclusions}
\label{sec:concl}
Our results indicate that large Hund interaction, $J_\mathrm{H}$, in the fully quantum Kondo-like models leads to the emergence of two types of quasiparticles in the system. The first ones resemble spinless fermions and constitute a band of nearly noninteracting quasiparticles. For a fully polarized system, the latter represents minority spin electrons, i.e., $\sigma=\downarrow$ electrons (being part of the doublon) in $\uparrow$-polarized system $T^z_\mathrm{tot}=SL+sL(2-n)$. The situation is more complicated for $T^z_\mathrm{tot}=0$ since one has to form a spinless quasiparticle out of two spinfull electrons. In the atomic limit, for $J_\mathrm{H}\gg t$ and $n>1$, the ground state is built out of doublons and three triplet projections, i.e., 
\begin{equation*}
|\mathrm{gs}\rangle_\mathrm{atomic}\simeq
\frac{1}{2}|\mathrm{D}_\uparrow\rangle+\frac{1}{2}|\mathrm{D}_\downarrow\rangle+
\frac{1}{2}|\mathrm{T}_0\rangle+\frac{1}{2}|\mathrm{T}_{\pm1}\rangle\,,
\end{equation*}
where
\begin{eqnarray*}
|\mathrm{D}_\uparrow\rangle&=&
\left|
\begin{matrix}
\uparrow\\
2
\end{matrix}
\right\rangle\,,\quad
|\mathrm{D}_\downarrow\rangle=
\left|
\begin{matrix}
\downarrow\\
2
\end{matrix}
\right\rangle
\,,\\
|\mathrm{T}_0\rangle&=&
\frac{1}{\sqrt{2}}\left|
\begin{matrix}
\uparrow\\
\downarrow
\end{matrix}
\right\rangle
+
\frac{1}{\sqrt{2}}\left|
\begin{matrix}
\downarrow\\
\uparrow
\end{matrix}
\right\rangle
\,\,,
\\
|\mathrm{T}_{\pm1}\rangle&=&
\frac{1}{\sqrt{2}}\left|
\begin{matrix}
\uparrow\\
\uparrow
\end{matrix}
\right\rangle
-
\frac{1}{\sqrt{2}}\left|
\begin{matrix}
\downarrow\\
\downarrow
\end{matrix}
\right\rangle
\,\,.
\end{eqnarray*}
Note that the doublons $|\mathrm{D}_\sigma\rangle$ are spinfull, i.e., each double occupancy in the itinerant band is accompanied by $\sigma=\uparrow$ or $\sigma=\downarrow$ spin in the localized band. Consequently, a natural candidate for low-energy (in proximity to the Fermi level) incoherent states that contribute to a spinless quasiparticle is the local singlet superposition of doublons
\begin{equation*}
|\mathrm{D}\rangle=
\frac{1}{\sqrt{2}}\left|
\begin{matrix}
\uparrow\\
2
\end{matrix}
\right\rangle
-
\frac{1}{\sqrt{2}}\left|
\begin{matrix}
\downarrow\\
2
\end{matrix}
\right\rangle\,.
\end{equation*}
In principle, one could also build the (local) singlet out of triplet projections (i.e., $|\mathrm{T}_0\rangle$ and $|\mathrm{T}_{\pm}\rangle$). Such a state would resemble the Affleck–Kennedy–Lieb–Tasaki (AKLT) state of $S=1$ Heisenberg model \cite{Affleck1987,Affleck1988} (the state realized for half-filling in the two-orbital Hubbard-Kanamori model \eqref{eq:ham2o} \cite{Mierzejewski2024}). However, the latter possesses topological properties, which we didn't observe in our investigations.

The second type of excitation forms a broad, incoherent band just below the Fermi level. Such excitations arise from the local triplets, which are part of the ground state in the atomic limit but are not part of the many-body state for a given polarization. For example, for $\uparrow$-polarized system, $T^z_\mathrm{tot}=SL+sL(2-n)$, the ground state is built out of doublons and $T^z_{\ell}=1$ local triplets, while the incoherent part has a projection on the local triplet with $T^z_{\ell}=0$, i.e., $|\mathrm{T}_0\rangle$. Note that such particles result from the quantum nature of localized spins, i.e., from the various local multiplets for given $n$.

Interestingly, these two types of excitations have vastly different behavior. One is spinless, i.e., carrying effective spin $S_\mathrm{eff}=0$, while the other $S_\mathrm{eff}=1$. The spinless quasiparticles exist above and below the Fermi level, and they mainly determine the properties of the system related to electron correlations, kinetic energy, and charge fluctuations (see App.~\ref{app:charge}). Our numerical results confirm such behavior, reproducing the noninteracting spinless solution perfectly. While the spinless particles are noninteracting, yielding a sharp $\cos$-like dispersion relation, a very broad and incoherent spectrum of the $S_\mathrm{eff}=1$-type indicates a strong interaction between them and/or with other degrees of freedom (e.g., with magnons). Due to the latter, the incoherent band is vital in understanding magnon decoherence.

For all considered electron densities, $n$, the dispersion of the spin excitations deviates strongly from the simple Holstein–Primakoff consideration for the nearest-neighbor exchange coupling $\propto [1-\cos(q)]$. In agreement with experimental investigation on manganites (which realize $J_\mathrm{H}|\mathbf{S}|\to\infty$ limit), we observe that magnons strongly decohere and/or change the dispersion towards the edge of the Brillouin zone, i.e., for $q\to\pi$. The strong damping of magnons can be explained as a consequence of their interaction with the Stoner-like continuum, which is built out of transitions between coherent spinless quasiparticles and incoherent excitations. It's important to note that such considerations go beyond the standard mean-field treatment of the Hund coupling. While the spinless quasiparticles appear already in $S\to\infty$ treatment of gK model \eqref{eq:hamkh} \cite{Yunoki1998}, the incoherent band of excitations is a consequence of the quantum nature of localized spin (i.e., $S=1/2$). Furthermore, it is worth noting that the excitations above the ferromagnetically ordered ground state don't depend strongly on the lattice dimensionality (at least within the Holstein–Primakoff treatment). The scenario presented in this work is a consequence of the various local multiplets present in the full quantum mechanical treatment of the problem, a phenomenon that is independent of lattice dimensionality. Consequently, our findings are relevant for a broad family of ferromagnetically ordered multiorbital compounds, especially displaying the OSMP properties.

Our results clearly show that magnon damping and mode softening in quantum double-exchange ferromagnets are present without the Jahn-Teller phonons (i.e., without any spin-lattice/orbit coupling in the model). The latter is the canonical explanation \cite{Khaliullin2000,Furukawa1999} of these phenomena for $J_\mathrm{H}|\mathbf{S}|\to\infty$ manganites. Although the Jahn-Teller distortion is necessary for the proper description of such compounds \cite{Hotta1999,Hotta2000,Yunoki2000,Dagotto2003,Pavarini2010}, our results give an alternative explanation for the origin of nontrivial spin dynamics. This is a remarkable result with important implications. There may be cases in materials where experimental features are believed to emerge from a combination of degrees of freedom that are not as active as assumed in the past.

\begin{acknowledgments}
A.M. and E.D. were supported by the US Department of Energy, Office of Science, Basic Energy Sciences, Materials Sciences and Engineering Division. G.A. was supported by the US Department of Energy, Office of Science, National Quantum Information Science Research Centers, Quantum Science Center. T.T. was supported by KAKENHI (Grant No. 24K00560) from the MEXT, Japan. J.H. acknowledges grant support by the National Science Centre (NCN), Poland, via Sonata BIS project no. 2023/50/E/ST3/00033. The calculations have been carried out using resources provided by the Wroclaw Centre for Networking and Supercomputing (\url{http://wcss.pl}). The \textsc{DMRG++} software developed in Oak Ridge National Laboratory was used for all calculations presented in this work. The code is available at \url{https://code.ornl.gov/gonzalo\_3/dmrgpp}. The input scripts for the \textsc{DMRG++} package are available at \url{https://bitbucket.org/herbrychjacek/corrwro/}.
\end{acknowledgments}

\appendix

\begin{figure*}[!htb]
\includegraphics[width=1.0\textwidth]{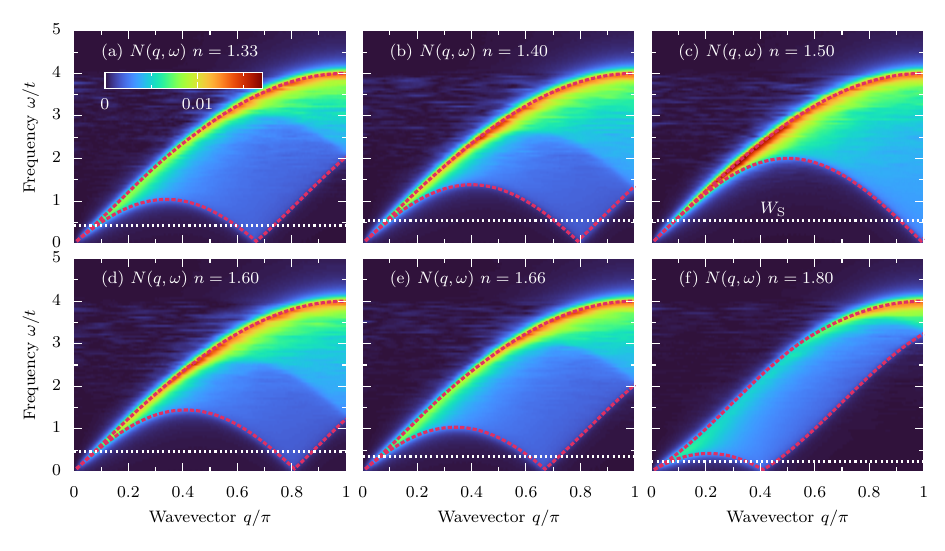}
\caption{Dynamical charge structure factor $N(q,\omega)$ of the gK model with $S=1/2$ localized spin in the $J_\mathrm{H}\gg t$ limit ($U/t=J_\mathrm{H}/t=20$, $T^z_\mathrm{tot}=0$ magnetization sector) for various electron doping levels, $n=1.33\,,1.40\,,1.50\,,1.60\,,1.66\,,1.80$ [panels (a) to (f), respectively]. The red dashed line represents the borders of the Stoner continuum calculated from noninteracting bands, Eq.~\eqref{eq:stonerff}. The white dashed line in all panels represents the span of the spin excitations $W_\mathrm{S}$. See Fig.~\ref{fig:width}(a). In all panels: $L=200$, $\delta\omega/t=5\cdot 10^{-3}$ and $\eta=2\delta\omega$.}
\label{fig:nqw}
\end{figure*}

\section{Charge dynamics}
\label{app:charge}

Here, we discuss the behavior of the dynamical charge structure factor defined as
\begin{equation}
N(q,\omega)=\frac{1}{L}\sum_{\ell} \mathrm{e}^{ i (\ell-L/2) q}\,
\langle\langle n_{\ell}n_{L/2} \rangle\rangle_{\omega}^{-}\,,
\label{eq:nqw}
\end{equation}
for the system parameters discussed in Fig.~\ref{fig:akw} and Fig.~\ref{fig:sqw} of the main text, i.e., for the generalized Kondo (gK) model with $S=1/2$ localized spins, $U/t=J_\mathrm{H}/t=20$, $L=200$ sites, and $T^z_\mathrm{tot}=0$ magnetization sector. Here $n_\ell=n_{\ell\uparrow}+n_{\ell\downarrow}$. In Fig.~\ref{fig:nqw}, we present $N(q,\omega)$ for various electron densities $n$. It is important to note that the total energy span of $N(q,\omega)$, and even the bottom of $N(q,\omega)$, lies much above the spin excitations bandwidth $W_\mathrm{S}$.

Our results indicate a perfect agreement between $N(q,\omega)$ obtained within the gK model in the $J_\mathrm{H}\gg t$ limit and the free-fermion solution. Specifically, for noninteracting spinless electrons, one can evaluate the charge structure factor $N(q,\omega)$ exactly \cite{Niemeijer1967,Muller1981}. Such calculations are equivalent to the Stoner continuum of the form
\begin{equation}
\omega_\mathrm{Sff}(q)=\omega_\mathrm{ff}(k_1)-\omega_\mathrm{ff}(k_2)\,,
\label{eq:stonerff}
\end{equation}
where $q=\mod(k_1+k_2,L)$, $k_1>k_\mathrm{F}$, and $k_2<k_\mathrm{F}$, and the free-fermion band $\omega_\mathrm{ff}(q)$, Eq.~\eqref{eq:ff}. Note that within Stoner-like considerations, one of the bands in~\eqref{eq:stonerff} represents $\sigma=\uparrow$ electrons, while the second band represents $\sigma=\downarrow$ electrons. The above perfect agreement between the noninteracting solution $\omega_\mathrm{Sff}(q)$ and full many-body calculations of $N(q,\omega)$ within the gK model~\eqref{eq:hamkh} in the $J_\mathrm{H}/t\gg 1$ limit indicates that the charge fluctuations are indifferent to the incoherent band of excitations.

\begin{figure*}[!htb]
\includegraphics[width=1.0\textwidth]{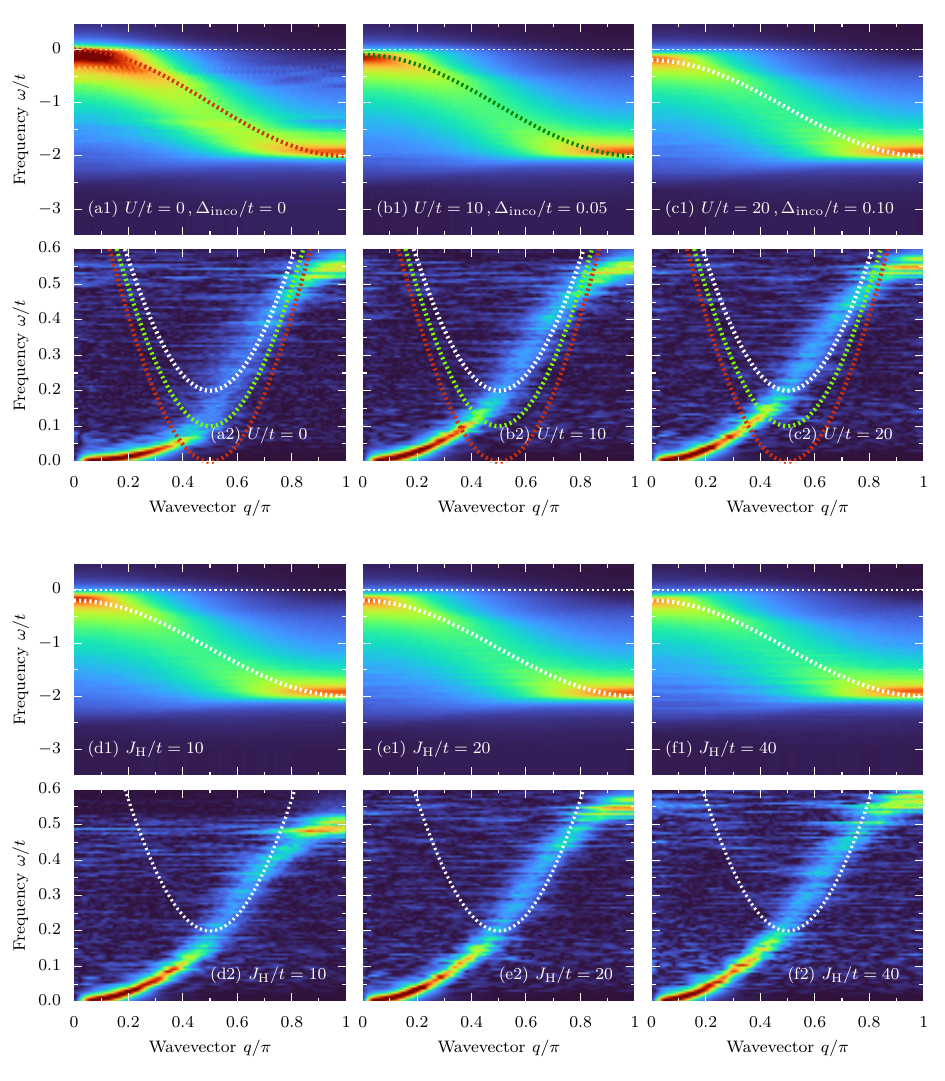}
\caption{Analysis of (a-c) the Hubbard interaction $U$ (for fixed $J_\mathrm{H}/t=20$) and (d-f) the Hund exchange (for fixed $U/t=20$) dependence of the single-particle spectral function $A(q,\omega)$ and the dynamical spin structure factor $S(q,\omega)$ of the gK model. (a1,b1,c1) Incoherent part of $A(q,\omega)$ as calculated for $U/t=0\,,10\,,20$, respectively. Dashed lines represents approximation $\omega_\mathrm{inco}(q)$, Eq.~\eqref{eq:incogap}, with $\Delta_\mathrm{inco}/t=0\,,0.05\,,0.1$. (a2,b2,c2) Spin excitations $S(q,\omega)$ are calculated for the corresponding values of $U$. Dashed lines (red, green, white) represent the bottom of the Stoner continuum $\omega_\mathrm{BS}(q)$ evaluated for $\Delta_\mathrm{inco}/t=0\,,0.05\,,0.1$, respectively. (d1-f1) Hund exchange $J_\mathrm{H}/t=10\,,20\,,40$ dependence of the incoherent part of $A(q,\omega)$ and (d2-f2) corresponding $S(q,\omega)$. Other parameters of the calculations: $A(q,\omega)$ data, panels (a1-f1), calculated with $\delta\omega/t=4\cdot 10^{-2}$ and $T^z_\mathrm{tot}=SL+sL(2-n)$. $S(q,\omega)$ data, panels (a2-f2), are calculated with $\delta\omega/t=6\cdot 10^{-3}$ and $T^z_\mathrm{tot}=0$. In all panels: $n=1.50$, $L=200$, and $\eta=2\delta\omega$.}
\label{fig:para}
\end{figure*}

\section{Hubbard and Hund interaction dependence}
\label{app:para}

In Sec.~\ref{sec:magston}, we demonstrated that the Hubbard interaction $U$ opens a small gap in the incoherent part of the single-particle spectral function $A(q, \omega)$. Here, we provide a detailed analysis of this phenomenon. Furthermore, we present additional results of $A(q, \omega)$ and the dynamical spin structure factor $S(q,\omega)$ for various values of the Hubbard $U$ and Hund $J_\mathrm{H}$ interaction.

As discussed in the main text, the role of the Hubbard interaction (even for large $U\gg t$) is minor. Only the incoherent part of the single-particle spectral function differs between different values of $U$. The detailed analysis of the latter is presented in Fig.~\ref{fig:para}. Upon increasing the value of $U$, one can observe that the large spectral weight of the incoherent part slowly shifts from $\omega=0$ for $U=0$, through $\omega/t\simeq0.05$ for $U/t=10$ to $\omega/t\simeq0.1$ at $U/t=20$ [see Fig.~\ref{fig:para}(a1,b1,c1)]. The latter yields an incoherent gap $\propto U$, i.e., $\Delta_\mathrm{inco}(U)/t=0,0.05,0.1$, for $U=0\,,10\,,20$, respectively. The Stoner continuum with the gap values corresponding to the position of the maximum spectral weight and the spin excitations $S(q,\omega)$ are presented in the second row of Fig.~\ref{fig:para}. We find that the region in which the magnons lose coherence for a given $U$ is better described by $\omega_\mathrm{inco}(q)$, Eq.~\eqref{eq:incogap}, and the corresponding bottom of the Stoner continuum $\omega_\mathrm{BS}(q)$ when the appropriate $\Delta_\mathrm{inco}(U)$ is included [see dashed lines in Fig.~\ref{fig:para}(a2,b2,c2)].

Finally, our analysis indicates that both $A(q,\omega)$ and $S(q,\omega)$ do not depend substantially on the values of the Hund exchange $J_\mathrm{H}$, provided that $J_\mathrm{H}\gg t$. In Fig.~\ref{fig:para}(c-f), we present results for $J_\mathrm{H}/t=10,20,40$. Specifically, the gap $\Delta_\mathrm{inco}$ does not change for all considered Hund values. Similarly, the region with a decreased magnon lifetime is the same for all considered $J_\mathrm{H}$.

\section{$S(q,\omega)$ fits details}
\label{app:fit}

In Sec.~\ref{sec:magsoft}, we have shown the analysis of the magnon dispersion relation obtained from the fits $\omega_\mathrm{fit}(q)$ to the maximum of $S(q,\omega)$ for given $q$ (i.e., to the data presented in Fig.~\ref{fig:sqw}). We have chosen $\omega_\mathrm{fit}(q)=a\tanh(b\,q^c)$ as a fit function, and the results of the procedure are given in Tab.~\ref{tab:tanfit}. Note that the functional form of $\omega_\mathrm{fit}(q)$ is arbitrary and "simple" polynomial fit $\omega_\mathrm{fit}(q)\propto\sum_i a_i\,q^i$ would yield similar results. Nevertheless, the $\tanh$ function is consistent across all considered electron densities $n$. Since in the inversion symmetric systems, one expects $q\to -q$ symmetry, we explicitly assume $\omega_\mathrm{fit}(2\pi-q)=\omega_\mathrm{fit}(q)$, i.e., we investigate only $0\leq q\leq\pi$.

\begin{table}[!htb]
\centering
\begin{tabular}{|c||c|c|c|c|c|c|c|c|}
\hline
$\,n\,$ & $\,1.25\,$ & $\,1.33\,$ & $\,1.40\,$ & $\,1.50\,$ & $\,1.60\,$ & $\,1.66\,$ & $\,1.75\,$ & $\,1.80\,$ \\ \hline\hline
$a$ & $0.30$ & $0.43$ & $0.55$ & $0.55$ & $0.47$ & $0.36$ & $0.23$ & $0.16$ \\ \hline
$b$ & $0.04$ & $0.06$ & $0.08$ & $0.14$ & $0.20$ & $0.30$ & $0.50$ & $0.60$ \\ \hline
$c$ & $3.0$  & $2.9$  & $2.8$  & $2.5$  & $2.2$  & $2.0$  & $2.0$  & $2.0$  \\ \hline
\end{tabular}
\caption{Fit parameters $a$, $b$, and $c$ of the \mbox{$\omega_\mathrm{fit}(q)=a\tanh(b\,q^c)$} function for various electron densities $n$.}
\label{tab:tanfit}
\end{table}

\bibliography{ferroosmp.bib}

\end{document}